\documentclass[superscriptaddress, noshowpacs, noshowkeys, twocolumn, prb]{revtex4}
\usepackage[italian, english]{babel}
\usepackage{amsmath,amssymb,amsfonts,latexsym}
\usepackage[pdftex]{graphicx}
\usepackage{color}
\pdfoutput=1

\usepackage[colorlinks,bookmarks=false,citecolor=blue,linkcolor=red,urlcolor=blue]{hyperref}

\newcommand\be{\begin{equation}}
\newcommand\bea{\begin{eqnarray}}
\newcommand\eea{\end{eqnarray}}
\newcommand\ee{\end{equation}}

\begin{document}

\title{Quantum quench from a thermal tensor state: \\ boundary effects and generalized Gibbs ensemble}
\author{Mario Collura}
\affiliation{Dipartimento di Fisica dell'Universit\`a di Pisa and INFN, 56127 Pisa, Italy}
\author{Dragi Karevski}
\affiliation{Institut Jean Lamour, dpt. P2M, Groupe de Physique Statistique, Universit\'e de Lorraine, 
CNRS, B.P. 70239, F-54506 Vandoeuvre-les-Nancy Cedex, France}

\date{\today}

\begin{abstract}
We consider a quantum quench in a non-interacting fermionic one-dimensional field-theory.
The system of size $L$ is initially prepared into two halves  $\mathcal{L}$ ($[-L/2,0]$) and 
$\mathcal{R}$ ($[0,L/2]$), each of them thermalized at two different 
temperatures, ${T_\mathcal{L}}$ and ${T_\mathcal{R}}$ respectively. 
At a given time the two halves are joined together by a local coupling and the whole system is left to evolve unitarily. 
For an infinitely extended system ($L\rightarrow \infty$), we show that the time evolution of the particle and energy densities is well described via a hydrodynamic approach which allows us to evaluate the correspondent stationary currents. 
We show, in such a case,  that the two-point correlation functions are deduced, at large times, from 
a simple non-equilibrium steady state. Otherwise, whenever the boundary conditions are retained 
(in a properly defined thermodynamic limit), any current is suppressed at large times, and the stationary
 state is described by a generalized Gibbs ensemble, 
 which is diagonal and depends only on the post-quench mode occupation.
\end{abstract}


\maketitle

\section{Introduction}

In the last few years there was a surprisingly growth on the theoretical study 
of the non-equilibrium properties of many-body quantum systems. 
This is chiefly due to the enhancement of the experimental techniques which
allow to manipulate with great precision trapped ultra-cold atomic gases,
without any significant coupling with the environment 
\cite{uc, kww06, tc07, tetal11, cetal12, getal11,shr12,rsb13,mhl12,fke13,mmk13}.
 Among all, of remarkable interest is the unitary dynamics 
of a quantum system initially prepared in a non-equilibrium state. 
In particular, an extremely important question is how to characterize the stationary state
to which the system should relax.
Indeed, depending on the integrability of the Hamiltonian governing the time evolution, the stationary
value of local observables can be described either by an effective Canonical ensemble or by a generalized
Gibbs ensemble (GGE), respectively for non-integrable and integrable systems 
(see Ref. \onlinecite{revq} for a review).
Many investigations have confirmed this scenario
\cite{rdyo07,rdo08,r09,rs12,wrdk13,mussardo10-13,cc06,c06,cdeo08,bs08,ce13,
CEF,scc09,f13,eef12,se12,fe13,csc13,kcc013,fcec13,bkc14,fagotti14,bhl12,krsm12, rsm12, mitra12,tm13,vlm13,ekmr13} 
and some effort have been also done in order to understand the role of the initial state in the construction of the GGE \cite{mussardo10-13,bkc14,fagotti14}.

In this respect, a very interesting non-equilibrium situation which is considered is a gas of atoms
 initially split into two different packets, each of them prepared at given initial temperature. 
The gas is then released  and left to evolve freely. 
A similar problem has already been addressed, using different analytical approaches, 
for systems defined on a lattice \cite{araki,ogata,aschbacher,pk07,dvbd13, kIm13}.
Nevertheless, those studies were focused on the non-equilibrium stationary state which is reached for an 
infinitely extended system. Indeed, such a non-equilibrium steady-state can be seen as
the state asymptotically realized from the initial state with its boundaries always connected to reservoirs 
at the initial temperatures \cite{mm11}. In other words, such infinitely extended system will always have, 
at each instant of time, two infinite portions (the far left and the far right playing the role of reservoirs) 
that remain equilibrated at the two initial temperatures. 
Between these two reservoirs, the (finite portion of the) system will show nonlocal properties 
reflecting the non-local structure of the non-equilibrium steady state (which is widely believed conjecture)
\cite{arrs98,ruelle98,jp02,kp09,prosen11,kps13,lm10,sm13}. 
Recently, by means of thermodynamic Bethe anstaz\cite{Zamo90}, 
the energy current in non-equilibrium steady states of integrable models of relativistic 
quantum field theory has been evaluated\cite{cacdh13} and interestingly, seems to be in disagreement 
with what has been numerically found in Ref. \onlinecite{kIm13} (at least within the numerical precision).

In this paper, we focus our attention on a gas of non-interacting spinless fermions either considering
the infinite-size limit or taking into account the effects of the boundaries.
 After preparing the system in a tensor thermal state, i.e. a tensor product of two different thermal density 
 matrices at two different temperature, we leave it to evolve with a non-interacting fermionic Hamiltonian.
 In particular, for an infinite system, by means of a hydrodynamic description \cite{akr08,cark12,wck13}, 
 we fully characterize the time-evolution of the particle density and energy density profiles. 
 From these results, we recover the Conformal Field Theory (CFT) predictions for the energy current
 flowing throughout the system \cite{bd12}. Furthermore, we inspect  the stationary state characterizing
 the infinitely extended system which does not agree with a {\it local} GGE description: 
 indeed, the local conserved charges appearing in the description of the non-equilibrium stationary state 
 combines in such a way that the resulting ``effective'' Hamiltonian is long-range interacting\cite{ogata}. 
 In this respect, it has been shown that such a non-locality may drastically affect the mutual information 
 between two adjacent subsystems\cite{ez13}.
 
 Notwithstanding, locality in the GGE is fully recovered whenever one preserves the effects of the boundaries. 
 We propose a genuine physical interpretation: the traveling
 particles, confined in a finite region, can reach the edges and, being reflected at different times, cause 
 a global dephasing giving rise to a true equilibrium state (no more currents) that is described by the GGE. In other words,
 we show that, in a properly defined thermodynamic (TD) limit, the reduced density matrix
 of any finite subsystem converges for long times to the GGE one. This implies that any measurable local 
 observable will converge to the GGE predictions.
 
 The paper is organized as follows. In Sec. \ref{model_postquench} we present the model and the
 post-quench Hamiltonian which governs the unitary evolution. In Sec. \ref{initial_system} we characterize
 the initial thermal tensor state. In Sec. \ref{quench_protocol} we study the time evolution of the two-point
  correlation function focusing our attention on the particle and energy density profiles. We introduce
  a hydrodynamic approach describing those quantities and we derive the corresponding particle and energy currents. 
  Finally, we also discuss the approach to the stationary values and we compare the two different scaling 
  regime (with or without boundaries) which lead to different asymptotic results. 
  In Sec. \ref{GGE} we finally show the structure of the GGE and we stress that all local observables 
  are described by the GGE in the fermionic momentum occupation numbers provided that the scaling limit
  is taken by considering the effects of the boundaries. Finally in Sec. \ref{Concl} we draw our conclusions.
\begin{figure*}[t]
\includegraphics[width=0.33\textwidth]{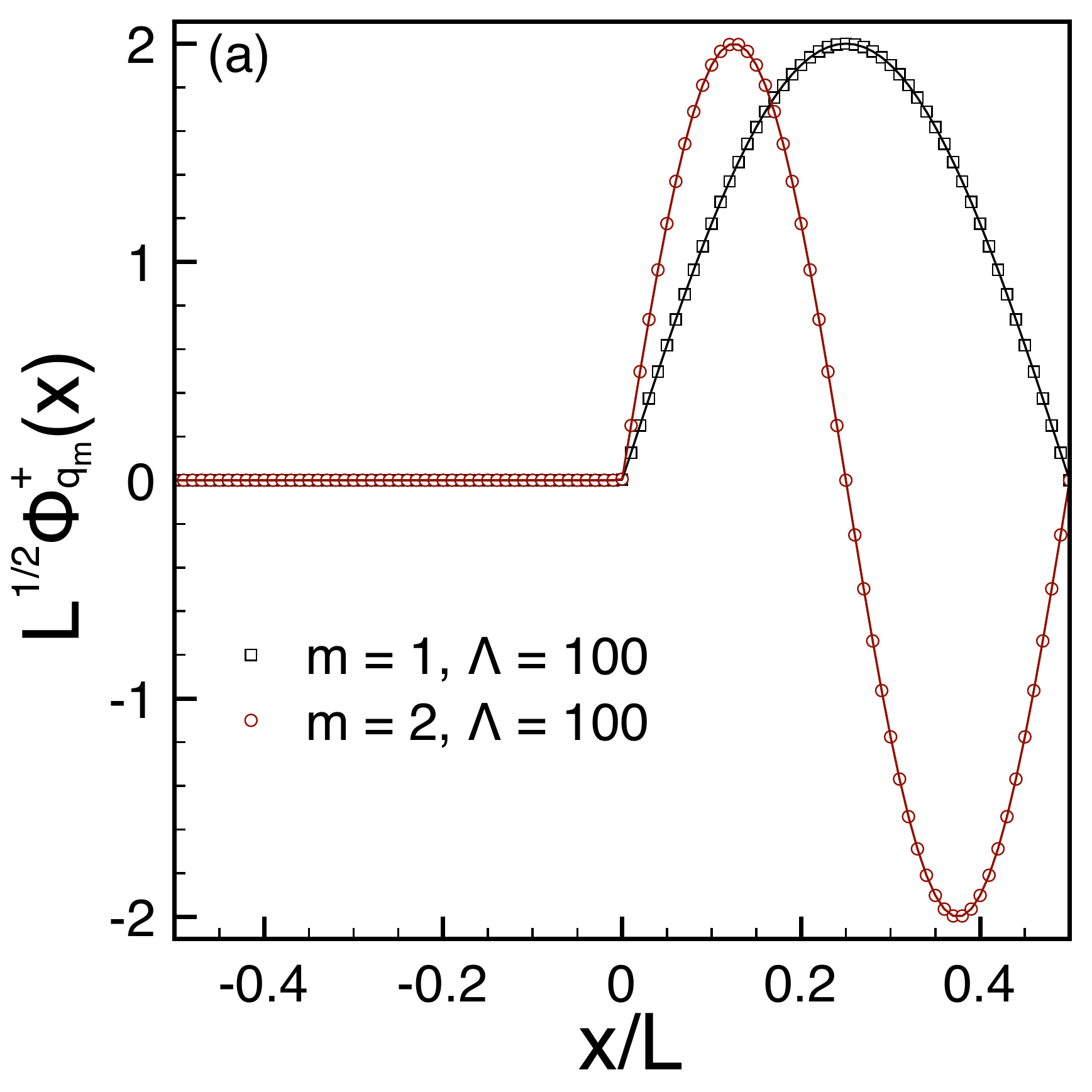}\includegraphics[width=0.33\textwidth]{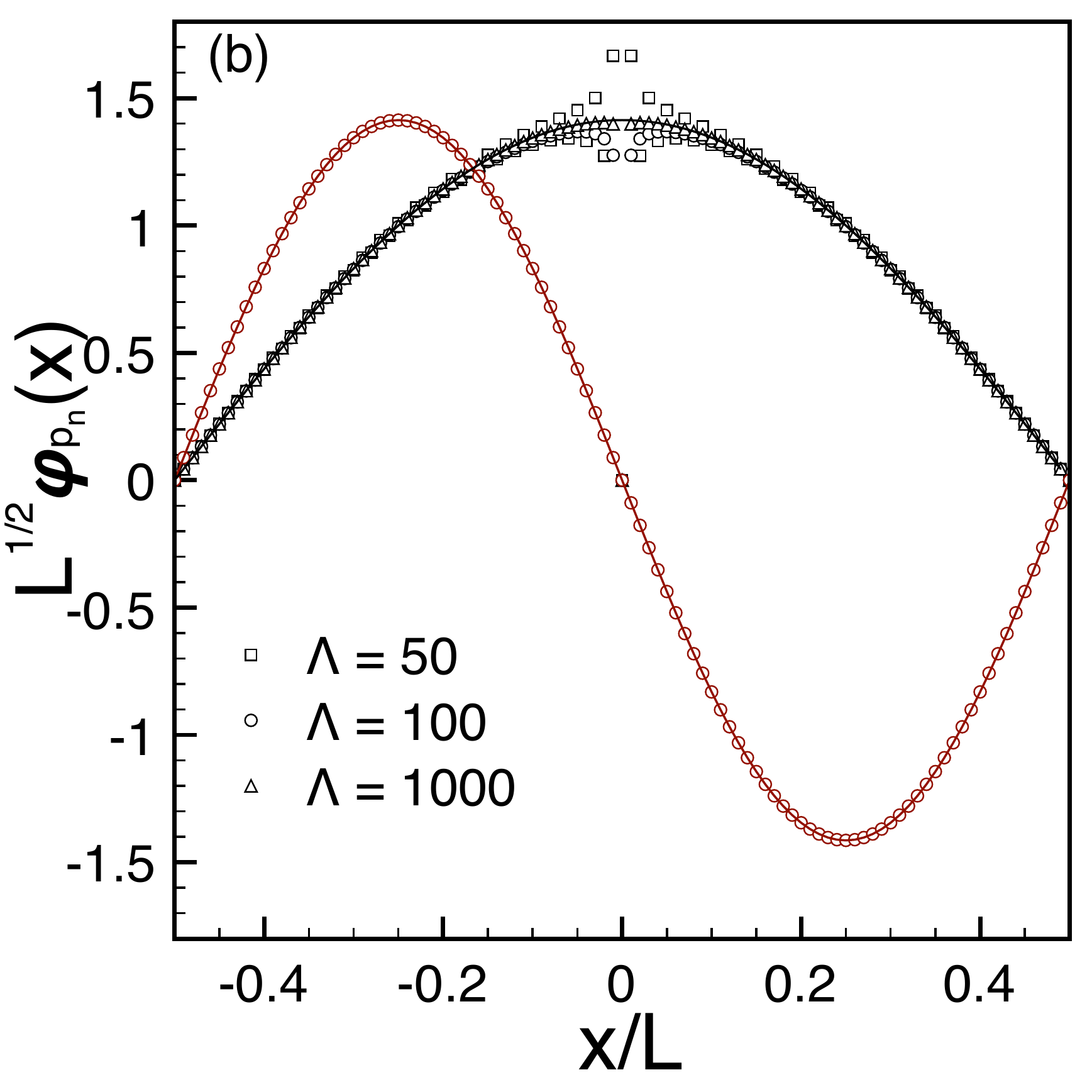}
\includegraphics[width=0.33\textwidth]{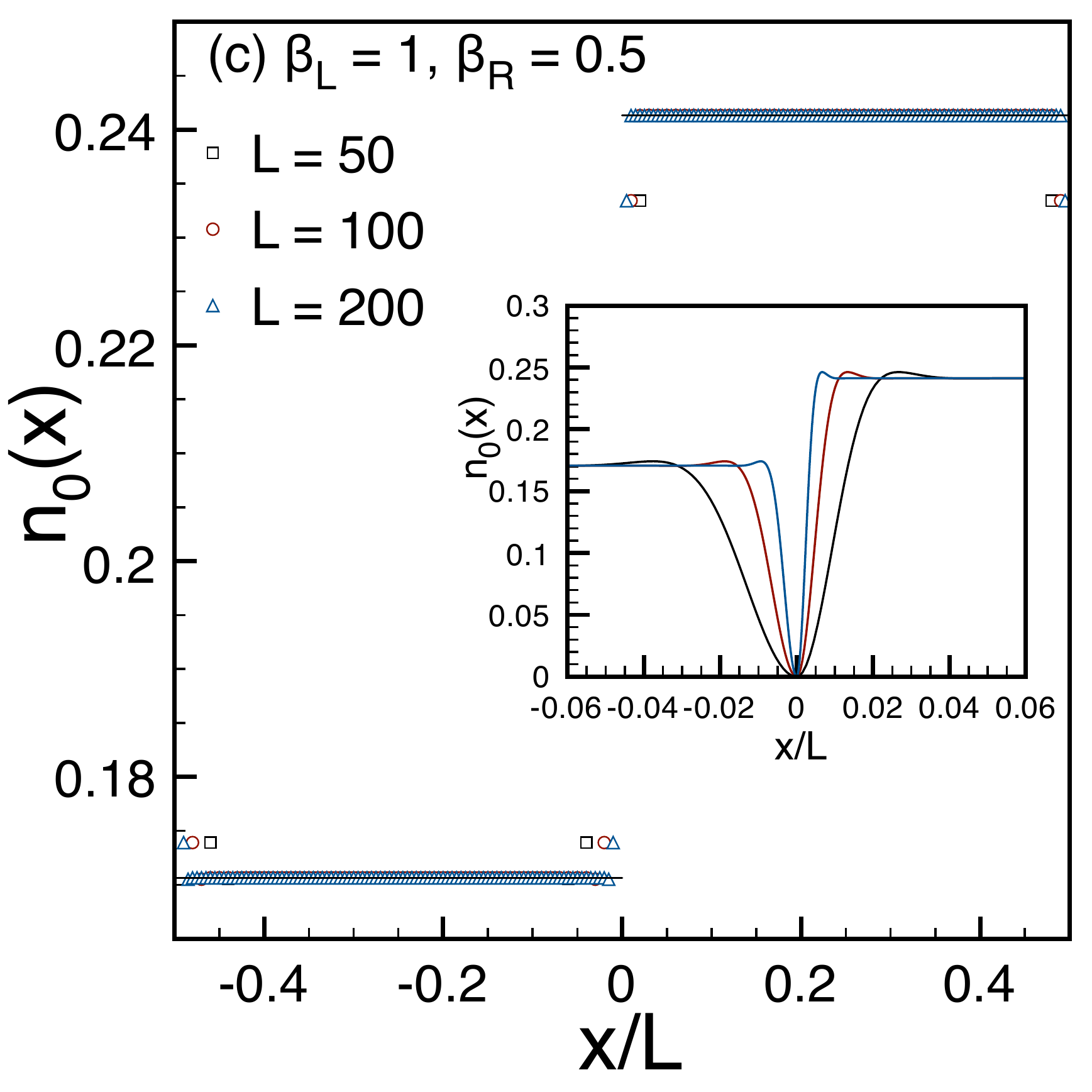}
\caption{(a) Pre-quench eigenfunctions $\phi^{+}_{q_m}(x)$ (full lines) compared with the series in
 Eq.(\ref{phi_to_varphi}) truncated at a given $\Lambda$ for the first two level $m=1,2$ (symbols). 
(b) The post-quench eigenfunctions $\varphi_{p_n}(x)$ (full lines) for $n=1,2$ compared with the series in
 Eq.(\ref{varphi_to_phi}) truncated at a given $\Lambda$ (symbols).
Notice how, for $n=1$ the sum approaches the eigenfunction apart from $x=0$,
wherein the series (\ref{varphi_to_phi}) is identically zero.
(c) Initial density profile for a system prepared with $\beta_{\mathcal{L}}=1$ and 
$\beta_{\mathcal{R}}=0.5$, and different sizes $L$. 
In the inset we zoomed the region close to the origin; notice how, the singular behavior in $x=0$ 
affects a region around the origin which shrinks as the system size $L$ increases.} 
\label{fig1}
\end{figure*}
  
\section{The model and the post-quench Hamiltonian}\label{model_postquench}

We consider a non-relativistic quantum-field theory describing non-interacting spinless fermions  in 
one spatial dimension. The system is defined in the symmetric interval $[-L/2,L/2]$ 
with open boundary conditions (OBC), and
where we introduced the system length $L$ in order to regularize infrared divergencies. 
At the end, we will be interested in the thermodynamic limit (TD limit) $L\to\infty$.
The Hamiltonian governing the unitary evolution of the system is given by
\be\label{H}
\hat H = \int_{-L/2}^{L/2} dx\,\partial_{x}\hat\Psi^{\dag}(x)  \partial_{x}\hat\Psi(x)\; ,
\ee
where the fields $\hat\Psi(x)$, $\hat\Psi^{\dag}(x)$ satisfy the canonical anti-commutation 
rules $\{\hat\Psi(x),\hat\Psi^{\dag}(y)\} = \delta(x-y)$. 
Introducing the normal free-fermionic operators
\bea\label{eta_to_Psi}
\hat\eta_{p_{n}} & = & \int_{-L/2}^{L/2} dx\, \varphi_{p_{n}}(x) \hat\Psi(x),\\
\varphi_{p_{n}}(x) & = & \sqrt{\frac{2}{L}}\sin \left[ p_{n} \left(x+ \frac{L}{2}\right) \right],\; p_{n} = \frac{\pi}{L} n \; ,
\eea
with $n\in\mathbb{N}$, the Hamiltonian (\ref{H}) is readily diagonalized
\be\label{H_diag}
\hat H = \sum_{n=0}^{\infty} p_{n}^{2} \hat{n}_{p_{n}},
\ee
with  $\hat{n}_{p_{n}}\equiv \hat\eta^{\dag}_{p_n}\hat\eta_{p_n}$ the fermionic mode occupation operator.
Notice that the Hamiltonian commutes with the total number of particles operator
\be
\hat{N} = \int_{-L/2}^{L/2}dx \, \hat\Psi^{\dag}(x)\hat\Psi(x) = \sum_{n=0}^{\infty}\hat{n}_{p_n},
\ee
therefore, $\hat H$ and $\hat N$ can be simultaneously diagonalized in the many-body Hilbert space. 
Since the excitation spectrum of the Hamiltonian is non-negative, 
the ground state without fixing the number of particles
is the vacuum state $|0\rangle$, such that $\hat\eta_{p_n}|0\rangle = 0,\;\forall n>0$.

\section{The initial system}\label{initial_system}
At time $t=0$ the system is divided into two halves $\mathcal{L}$ ($x<0$) and $\mathcal{R}$ ($x>0$). 
Therefore the two subsystems are initially uncorrelated and the total initial Hamiltonian is the direct sum
of two independent Hamiltonians which can be diagonalized separately. 
Indeed, the Hamiltonian describing a subsystem is given by Eq.(\ref{H}) 
where the integration domain is reduced to a half interval: $[0,L/2]$ or $[-L/2,0]$ 
for the right and left subsystems respectively. 

The normalized eigenfunctions, building-up 
the one-particle Hilbert space of the semi-intervals, are given by
\be
\phi^{\pm}_{q_{m}}(x) = \frac{2}{\sqrt{L}}\sin\left( q_{m} x \right) \theta(\pm x),\; q_{m} = \frac{2\pi}{L} m, \, m\in\mathbb{N},
\ee
where $\theta(x)$ is the Heaviside function such that $\theta(x)=1$ for $x>0$, and zero otherwise.
Thanks to these functions the half-interval Hamiltonians are straightforwardly diagonalized. For example,
in the sub-interval $[0,L/2]$ one has
\be\label{H_0}
\hat H^{+}_{0} = \int_{0}^{L/2} dx\,\partial_{x}\hat\Psi^{\dag}(x)  \partial_{x}\hat\Psi(x) 
=  \sum_{m=0}^{\infty} q_{m}^{2} \hat\xi_{q_m}^{\dag}\hat\xi_{q_m},
\ee
where the fermionic fields $\Psi$, $\Psi^{\dagger}$ are related to the diagonal operators via
\bea
\hat\Psi(x) = \sum_{m=0}^{\infty} \phi^{+}_{q_m}(x) \hat\xi_{q_m},\quad x>0.
\eea
Similar arguments are valid for the negative half-interval. 

\subsection{The overlap}
It is  useful  to express the initial one-particle eigenfunctions 
$\phi^{\pm}_{q_m}(x)$ in terms of the post-quench one-particle eigenfunctions $\varphi_{p_n}(x)$, i.e.
\be\label{phi_to_varphi}
\phi^{\pm}_{q_m}(x) = \sum_{n=0}^{\infty} A^{\pm}_{n,m} \varphi_{p_n}(x),
\ee
where we have explicitly introduced the overlap
\be
A^{\pm}_{n,m}  = \int_{-L/2}^{L/2} dx \, \varphi_{p_n}(x) \phi^{\pm}_{q_m}(x).
\ee
Notice that Eq. (\ref{phi_to_varphi}) is exact since the post-quench one-particle eigenstates are a 
complete basis in the symmetric interval $[-L/2,L/2]$. Otherwise, in order to invert such a relation, and to 
rewrite $\varphi_{p_n}(x)$ in terms of the pre-quench eigenstates, one needs a linear combination of 
both $\phi^{-}_{q_m}(x)$ and $\phi^{+}_{q_m}(x)$,
\be\label{varphi_to_phi}
\varphi_{p_n}(x) = \sum_{m=0}^{\infty} [ A^{-}_{n,m} \phi^{-}_{q_m}(x) + A^{+}_{n,m} \phi^{+}_{q_m}(x) ].
\ee
Nevertheless, for $n$ odd, the series in Eq.(\ref{varphi_to_phi}) is not absolutely convergent in $[-L/2,L/2]$ 
due to the anomalous behavior in $x=0$. However, such anomaly can be neglected in the thermodynamic limit
since $x=0$ is a subset of measure zero in $\mathbb{R}$. The overlap can be explicitly evaluated, giving
\be
A^{\pm}_{n,m} = \left\{\begin{array}{cc}
\pm \frac{4\sqrt{2} m \sin(n \pi /2) }{\pi(4m^2-n^2)} & \mathrm{for}\; n\;\mathrm{odd}, \\
 & \\
 \frac{(-1)^m}{\sqrt{2}}\delta_{n,2m} & \mathrm{for}\; n\;\mathrm{even}.
\end{array}\right. 
\ee

In Figure \ref{fig1} we show how the series in Eq.(\ref{phi_to_varphi}) and Eq.(\ref{varphi_to_phi}) 
reproduce the correct post(pre)-quench one-particle eigenfunctions.

\begin{figure*}[t!]
\includegraphics[width=0.33\textwidth]{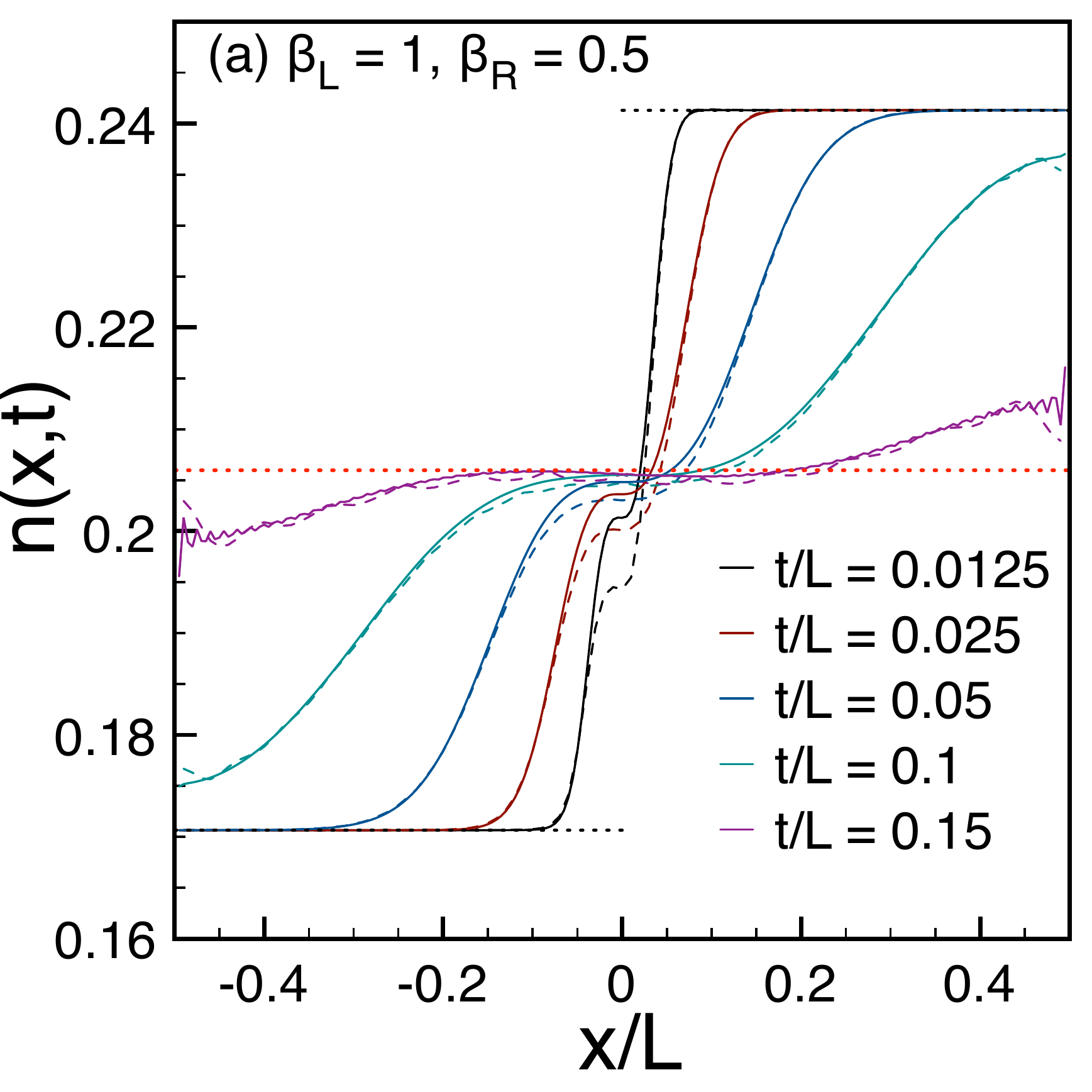}\includegraphics[width=0.33\textwidth]{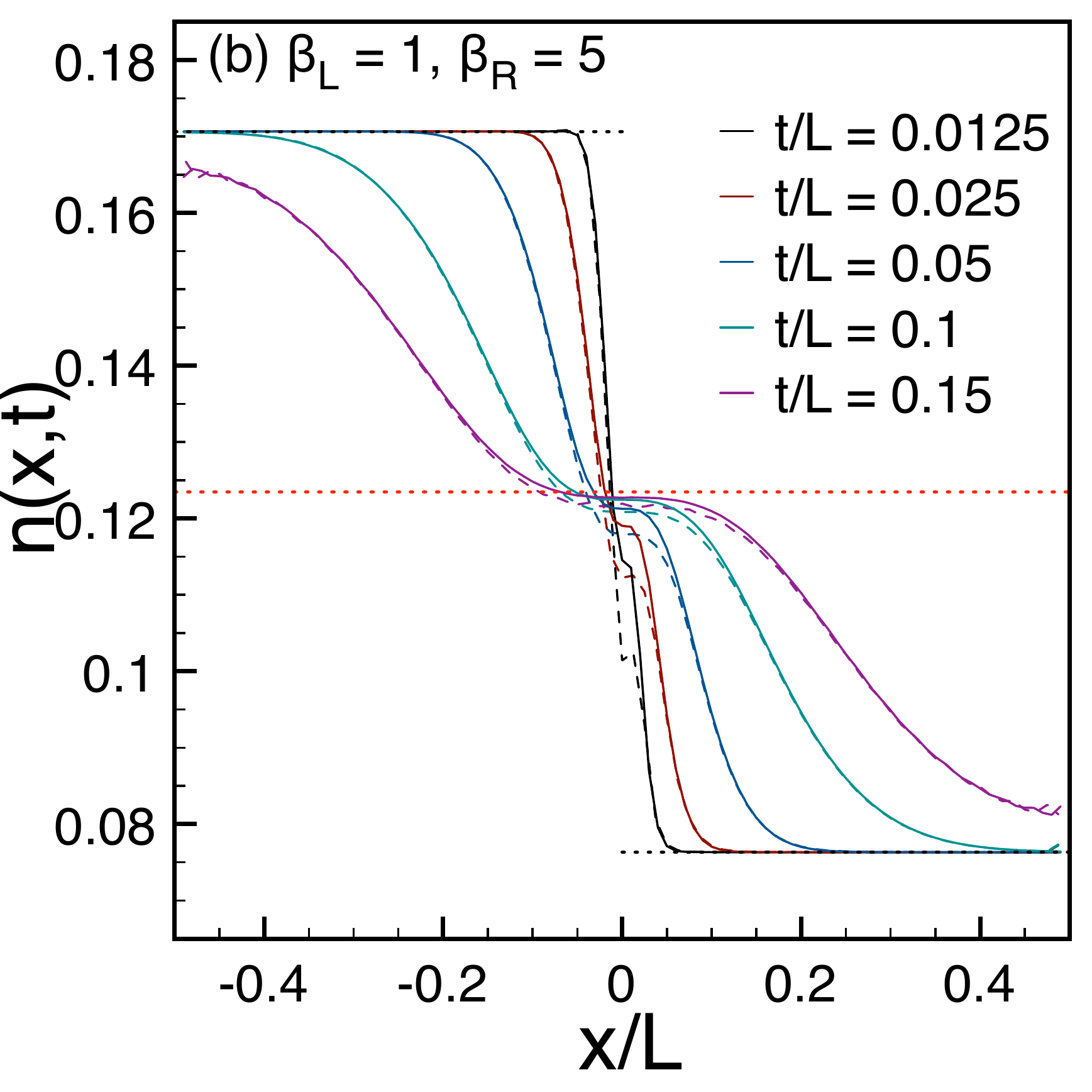}
\includegraphics[width=0.33\textwidth]{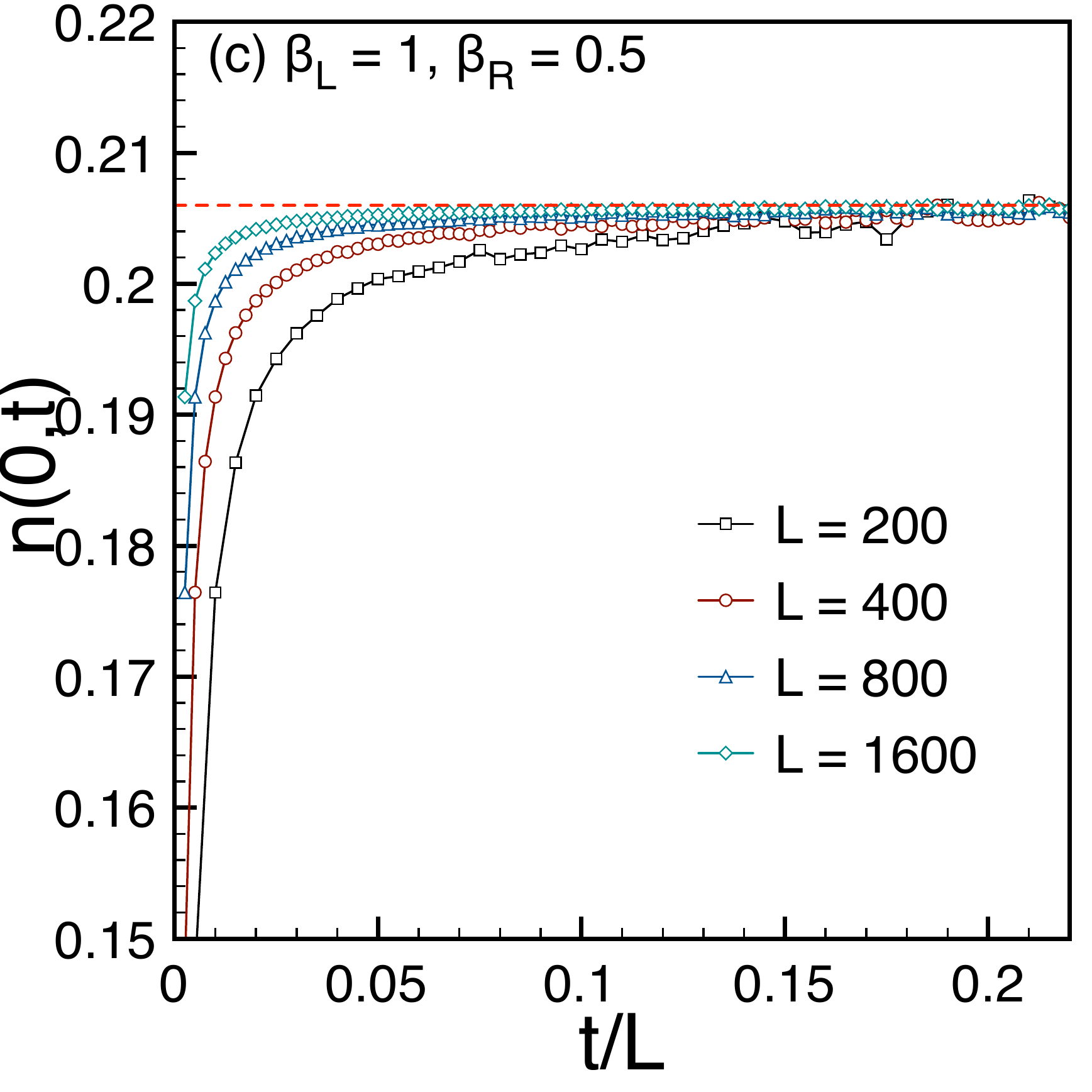}
\caption{Particle density profiles vs rescaled space $x/L$ at different rescaled times $t/L$ for a system initially 
prepared with $\beta_{\mathcal{L}}=1$ and $\beta_{\mathcal{R}}=0.5$ (a) and $\beta_{\mathcal{R}}=5$ (b). 
Dashed lines correspond to $L=400$, full lines to $L=1000$. 
Dotted lines represent the initial densities (black dotted) and the stationary density (red dotted) in the TD limit.
(c) Evolution of the local particle density at $x=0$ for different sizes $L$. 
Notice how, in terms of the rescaled time $t/L$,
the curves approach a step-like function as $L$ increases.} 
\label{fig_nxt}
\end{figure*}

\subsection{Thermal tensor state}
The initial state is constructed as a tensor product of two thermal density matrices at two different temperatures, 
i.e. $\hat\rho_{0} = \hat\varrho_{-}(\beta_{\mathcal{L}})\otimes\hat\varrho_{+}(\beta_{\mathcal{R}})$, 
where $\hat\varrho_{\pm}(\beta) = Z^{-1} \exp (-\beta \hat H_{0}^{\pm})$. 
This means that the two spatial regions ($\mathcal{L}$ and $\mathcal{R}$) are initially uncorrelated. Furthermore, 
such a state is quadratic in the local field operators and, therefore, the Wick's theorem applies.

Thanks to this fact, the initial two-point correlation function 
$C_{0}(x,y)\equiv \langle \hat\Psi^{\dag}(x)\hat\Psi(y) \rangle_0$ splits into two separate terms
\be
C_{0}(x,y)  =  C_{\mathcal{L}}(x,y)\theta(-x)\theta(-y) + C_{\mathcal{R}}(x,y)\theta(x)\theta(y)
\ee
with
\bea\label{C0_disc}
C_{\mathcal{L/R}}(x,y) & = & \frac{4}{L}\sum_{m=0}^{\infty}\frac{\sin(q_m x)\sin(q_m y)}{1+\exp(\beta_{\mathcal{L/R}} q_{m}^{2})}\\
& = & \frac{2}{L}\sum_{m=0}^{\infty}\frac{\cos[q_m (x-y)]}{1+\exp(\beta_{\mathcal{L/R}} q_{m}^{2})}\\
& - & \frac{2}{L}\sum_{m=0}^{\infty}\frac{\cos[q_m (x+y)]}{1+\exp(\beta_{\mathcal{L/R}} q_{m}^{2})}.
\eea
Notice that the initial correlation function depends both on $x-y$ and $x+y$ due to the fact that
the initial state breaks the translational invariance, even in the thermodynamic limit. 
Indeed, taking the TD limit, thanks to the parity of the fermionic distribution, we can rewrite the latter as
\bea\label{C0_cont}
C_{\mathcal{L/R}}(x,y) & = & \beta_{\mathcal{L/R}}^{-1/2} \mathcal{F} \left[(x-y) \beta_{\mathcal{L/R}}^{-1/2} \right] \\
& - & \beta_{\mathcal{L/R}}^{-1/2} \mathcal{F} \left[(x+y) \beta_{\mathcal{L/R}}^{-1/2} \right],
\eea
where we introduced the Fourier transform
\be\label{F(z)}
\mathcal{F}(z) = \int_{-\infty}^{\infty} \frac{dq}{2\pi} \frac{\exp(i z q)}{1+\exp(q^2)}.
\ee
Using the Taylor's series expansion $(1+z)^{-1} = -\sum_{n=1}^{\infty}(-1)^{n}z^{-n}$, we have
\be
\mathcal{F} (z) = -\frac{1}{2\sqrt{\pi}}\sum_{n=1}^{\infty}\frac{(-1)^{n}\exp[-z^{2}/(4n)]}{\sqrt{n}}.
\ee
Integrating Eq.(\ref{C0_disc}) with $y=x$, and using the orthogonality of the one-particle 
eigenfunctions, one immediately obtains the left (right) number of particles
\bea
N_{\mathcal{L/R}} & = & \sum_{m=0}^{\infty} \frac{1}{1+\exp(\beta_{\mathcal{L/R}} q_{m}^{2})}\\ 
& = & \frac{L}{\sqrt{\beta_{\mathcal{L/R}}}} \int_{0}^{\infty} \frac{dq}{2\pi} \frac{1}{1+\exp(q^{2})}\\
& = & L \frac{(1-\sqrt{2})\zeta(1/2)}{4\sqrt{\pi}\sqrt{\beta_{\mathcal{L/R}}}}
\eea
where $\zeta(z) = \sum_{k=1}^{\infty} k^{-z}$ is the Reimann zeta function. 
Notice that the number of particle is, as it should, diverging in the TD limit making finite the 
initial left and right densities $n_{\mathcal{L/R}} \equiv 2\,N_{\mathcal{L/R}}/L$.  Moreover, in the TD limit
the left and right initial densities are uniform. In Figure \ref{fig1} we plot the numerical evaluated initial
densities for different system sizes.

In the same way, we can evaluate the left (right) total energy in the TD limit. 
Indeed, from Eq.(\ref{H_0}) one immediately has
\bea
E_{\mathcal{L/R}} & = &\sum_{m=0}^{\infty} \frac{q_{m}^2}{1+\exp(\beta_{\mathcal{L/R}} q_{m}^{2})}\\ 
 & = & \frac{L}{ \beta_{\mathcal{L/R}}^{3/2} } \int_{0}^{\infty} \frac{dq}{2\pi} \frac{q^2}{1+\exp(q^{2})}\\
 & = & L \frac{(2-\sqrt{2})\zeta(3/2)}{16 \sqrt{\pi} \, \beta_{\mathcal{L/R}}^{3/2} },
\eea
which gives rise to finite energy densities $\mathcal{E}_{\mathcal{L/R}} \equiv 2 E_{\mathcal{L/R}} /L$. 

\begin{figure*}[t!]
\includegraphics[width=0.5\textwidth]{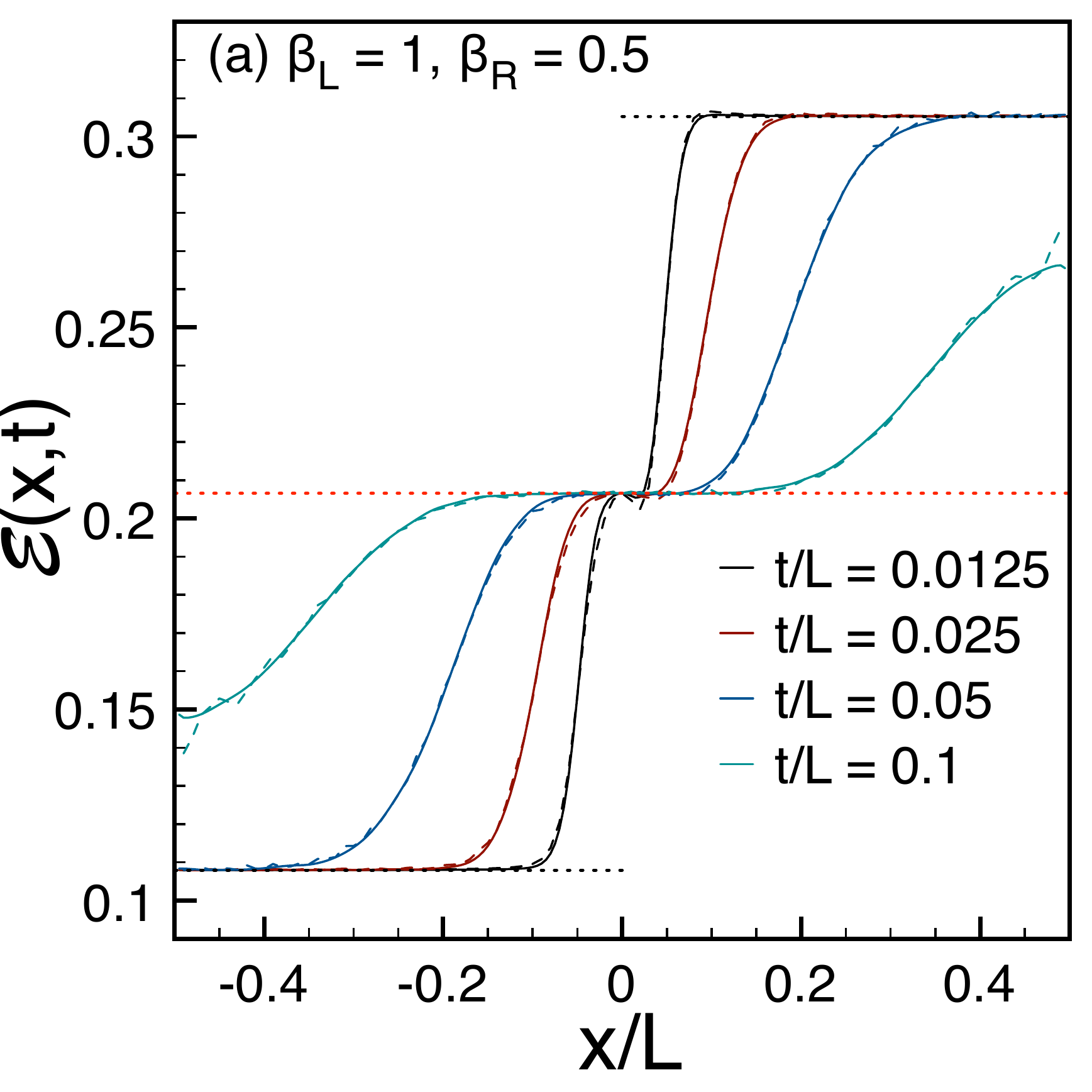}\includegraphics[width=0.5\textwidth]{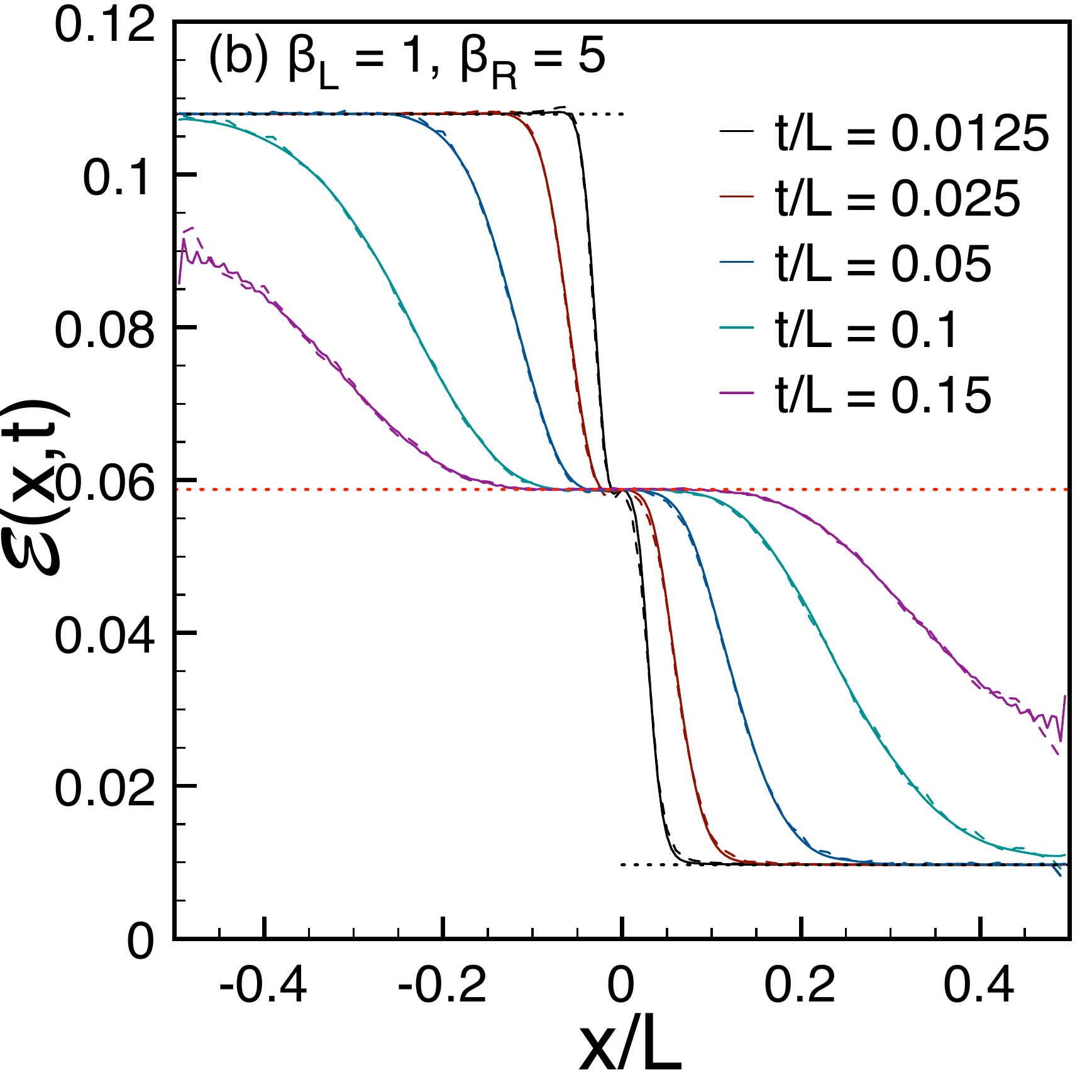}
\caption{Energy density profiles vs rescaled space $x/L$ at different rescaled times $t/L$ for a system initially 
prepared with $\beta_{\mathcal{L}}=1$ and $\beta_{\mathcal{R}}=0.5$ (a) and $\beta_{\mathcal{R}}=5$ (b). 
Dashed lines correspond to $L=400$, full lines to $L=1000$. 
Dotted lines represent the initial densities (black dotted) and the stationary density (red dotted) in the TD limit.} 
\label{fig_Ext}
\end{figure*}

\section{The quench protocol}\label{quench_protocol}
In this section we analyze the out-off-equilibrium dynamics after putting in contact the two 
halves $\mathcal{L}$ and $\mathcal{R}$. In other words, the initial state $\hat\rho_{0}$ evolves  unitarily 
according to the post-quench Hamiltonian $\hat H$. In particular, since the Hamiltonian is quadratic in the
fermionic operators, the evolved state keeps its gaussian character and the Wick's theorem still applies. 
Consequently, all observables are derived from the two-point correlation function. 
Therefore, we focus our attention on the time-evolution of the two-point correlation function which is given by
\be\label{Corr_t}
C(x,y;t) = \sum_{n,m = 0}^{\infty} \varphi_{p_n}(x)\varphi_{p_m}(y) \langle \hat\eta^{\dag}_{p_n} \hat\eta_{p_m} \rangle_{t}, 
\ee
where the evolution of the diagonal operators is trivially given by
\be\label{etaeta_t}
\langle \hat\eta^{\dag}_{p_n} \hat\eta_{p_m} \rangle_{t} = \exp[i (p_n^2-p_m^2)t] 
\langle \hat\eta^{\dag}_{p_n} \hat\eta_{p_m} \rangle_{0}.
\ee
Using the inverse transformation of Eq. (\ref{eta_to_Psi}), the initial correlation function of the post-quench 
fermionic operators can be rewritten in terms of $C_{0}(x,y)$
\be
\langle \hat\eta^{\dag}_{p_n} \hat\eta_{p_m} \rangle_{0} = 
\iint dz dw\, \varphi_{p_n}(z)\varphi_{p_m}(w) C_{0}(z,w).
\ee
Using the overlap between the pre-quench and the post-quench
one-particle eigenfunctions one obtains
\be\label{etaeta_0}
\langle \hat\eta^{\dag}_{p_n} \hat\eta_{p_m} \rangle_{0}  =
\sum_{l=0}^{\infty} \left [ \frac{A^{-}_{n,l}A^{-}_{m,l}}{1+\exp(\beta_{\mathcal{L}} q_{l}^{2})}
+ \frac{A^{+}_{n,l}A^{+}_{m,l}}{1+\exp(\beta_{\mathcal{R}} q_{l}^{2})} \right].
\ee
Finally, using the latter equation, we can rewrite the time evolved correlation function as
\be
C(x,y;t) = \sum_{l=0}^{\infty} \left[ \frac{\phi^{-}_{q_l}(x,t)^{*}\phi^{-}_{q_l}(y,t)}{1+\exp(\beta_{\mathcal{L}} q_{l}^{2})} 
+  \frac{\phi^{+}_{q_l}(x,t)^{*}\phi^{+}_{q_l}(y,t)}{1+\exp(\beta_{\mathcal{R}} q_{l}^{2})} \right],
\ee
where we introduced the time-evolved one particle eigenfunctions
\be
\phi^{\pm}_{q_m}(x,t) = \sum_{n=0}^{\infty}A^{\pm}_{n,m}\varphi_{p_n}(x) \mathrm{e}^{-i p_n^2 t}.
\ee

We numerically evaluated the time-dependent particle density $n(x,t)\equiv C(x,x;t)$ and we report in 
Figure \ref{fig_nxt} the profile at different times for a system of sizes $L=400,1000$.
We can see how the density equilibrate toward the stationary value: a causality zone, 
with homogeneous density, propagates starting from the origin 
and determines, at each instant of time, an equilibration region $[-x^{*}(t), x^{*}(t)]$.
Notice, however that, for finite systems, the singular behavior in $x=0$ 
affects the value of the density in such region, which displays size effects. 
However, whenever the TD limit is considered, the plateau in the density profile around $x=0$ takes a constant 
value converging as the size is increased to the large-time stationary value (see Figure \ref{fig_nxt} ). 

Furthermore, we also numerically evaluated the energy density profile 
$\mathcal{E}(x,t) \equiv \langle \partial_{x}\hat\Psi^{\dag}(x)\partial_{x}\hat\Psi(x) \rangle_{t}$ which 
can be easily shown to be equal to $\partial^{2}_{xy}C(x,y;t)|_{y=x}$. 
In Figure \ref{fig_Ext} we plot the energy profiles for different system sizes and different initial temperatures. 
The same considerations that was done for the particle density applies also in this case. Nevertheless, in 
this case, thanks to the fact that the energy density depends on the derivative of the one-particle eigenfunctions,
the anomalous behavior in the neighborhood of  $x=0$ (which is evident in the particle density)
 is smoothed out.   
\begin{figure}[t!]
\includegraphics[width=0.245\textwidth]{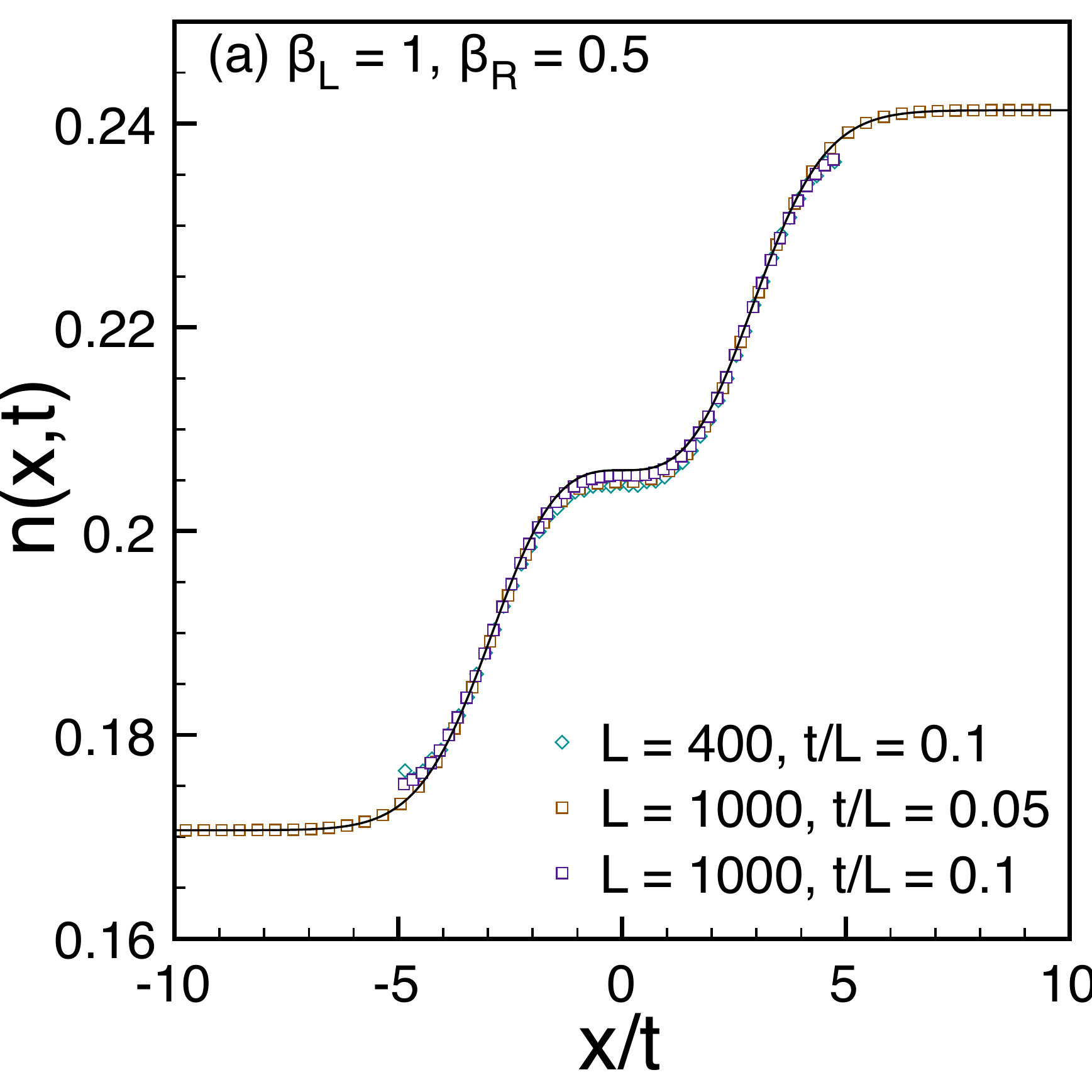}\includegraphics[width=0.245\textwidth]{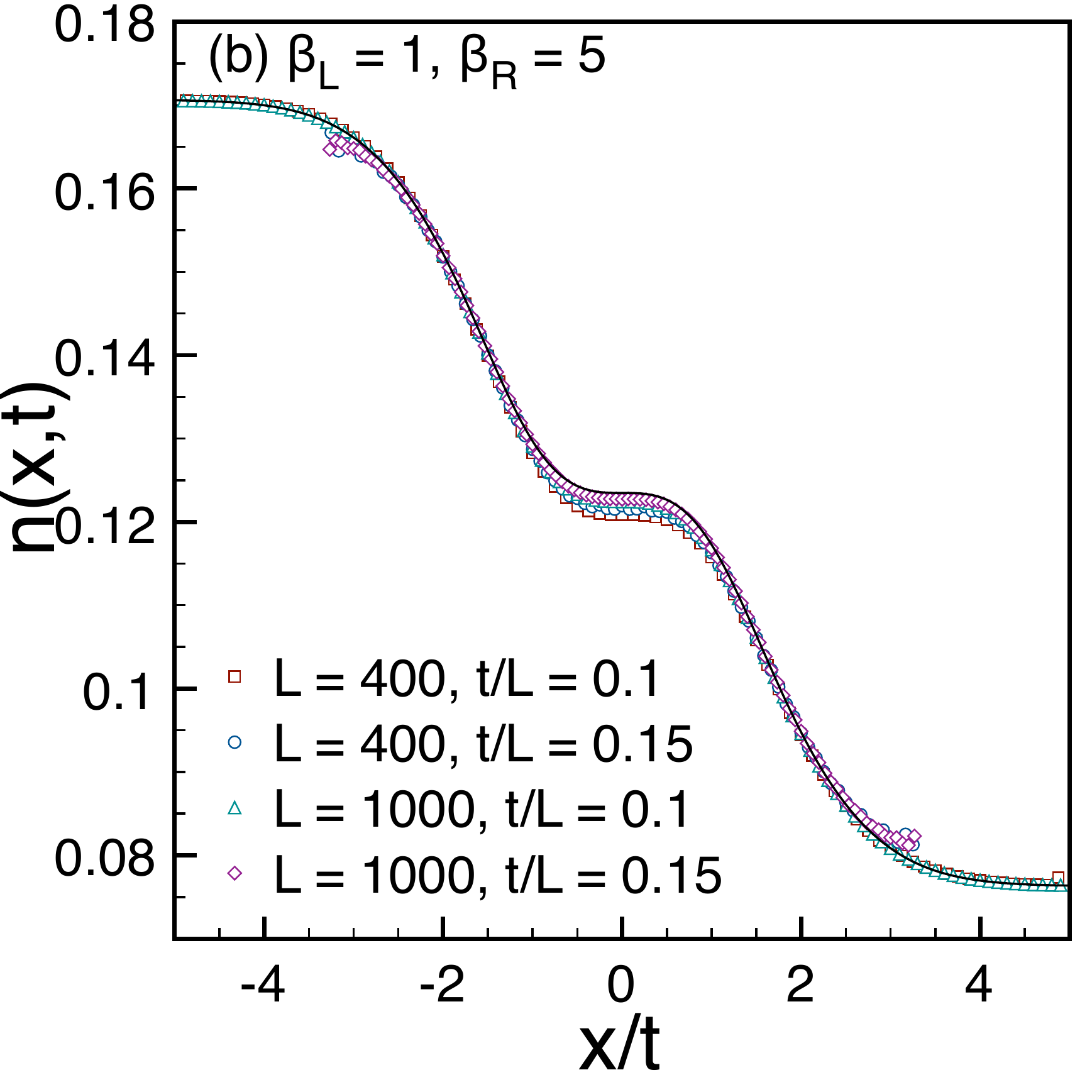}
\includegraphics[width=0.245\textwidth]{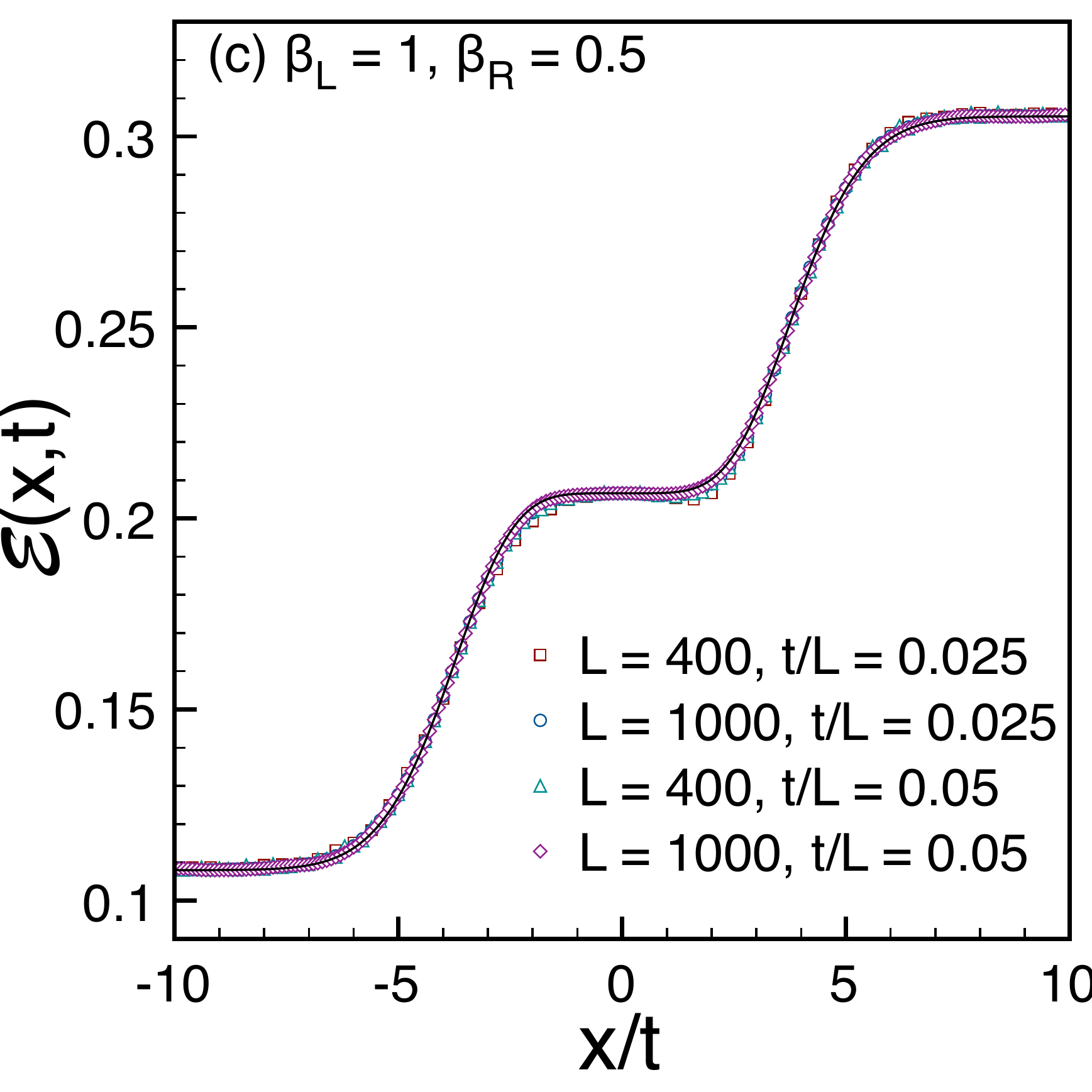}\includegraphics[width=0.245\textwidth]{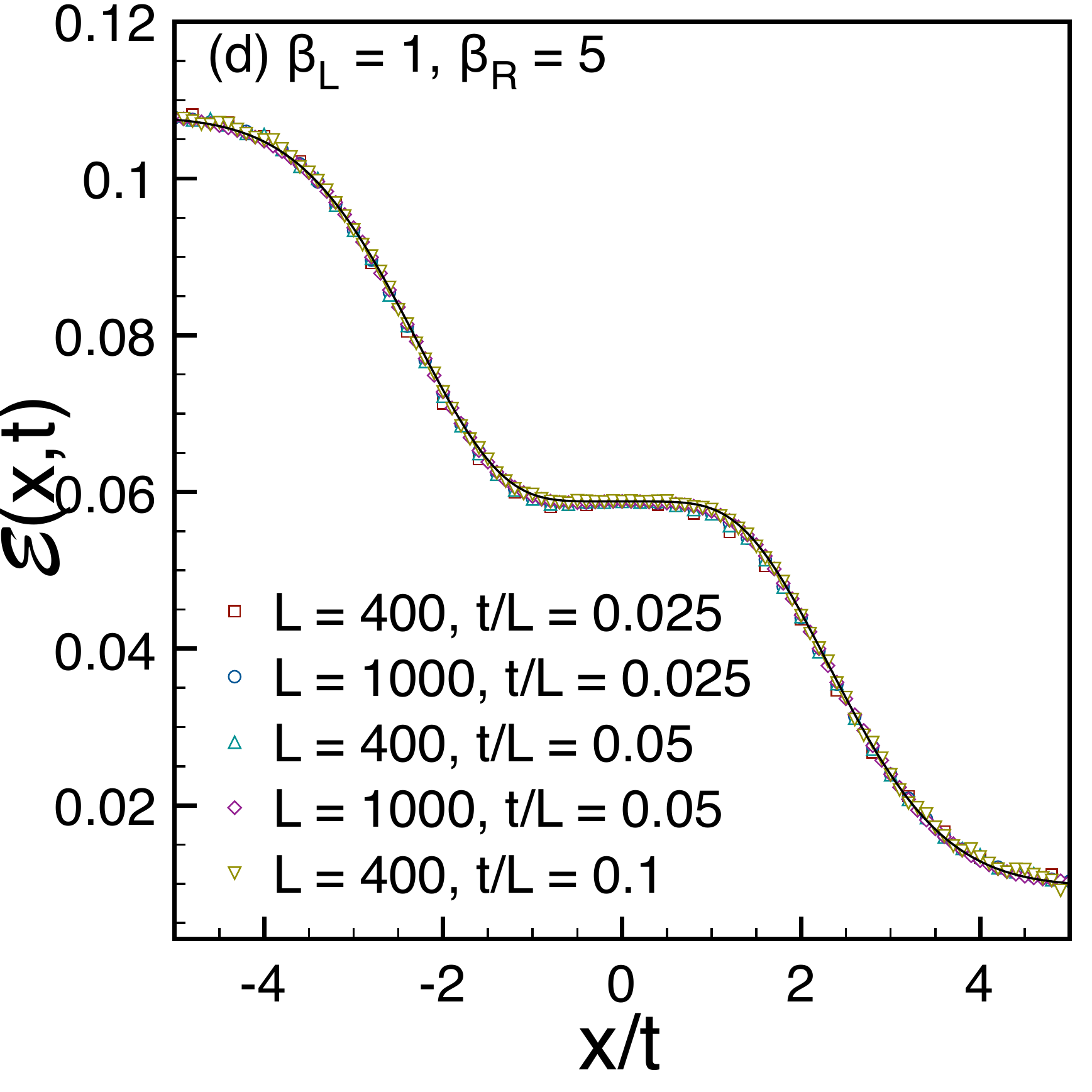}
\caption{(a,b) Numerically evaluated particle density profile vs the scaling variable $x/t$ for different sizes $L$
and rescaled times $t/L$ (symbols). The full lines are the analytic result (\ref{nxt_scaling}). (c,d) Numerically
 evaluated energy density profile vs the scaling variable $x/t$ for different sizes $L$
 and rescaled times $t/L$ (symbols). The full lines are the analytic result (\ref{Ext_scaling}) } 
\label{fig_nE_scaling}
\end{figure}
 
 \subsection{Hydrodynamic description of the time-dependent profile of local observables}
 Taking into account all the considerations of the previous paragraphs, we argue that the dynamics of the
 particle-density profile as well as of the energy-density profile can be fully characterized, in the TD limit, 
 by a semi-classical description. We want to stress here that such a description, by construction, 
 can not reproduce the dynamics of the correlations. The idea underlying the semi-classical approach was
 already proposed in Ref. \onlinecite{akr08,cark12,wck13} and is essentially based on the fact that the quantum 
 dynamics of the local densities is well captured by a hydrodynamic description 
 in the phase-space $(p,x)$ of the correspondent classical coarse-grained densities.
 
 Indeed, if one considers, for example, the particle density, we can associate to each phase-space point
 a local initial packet of particles $n_{0}(p,x) dp dx$. Then, each of them evolves following a classical trajectory 
 $x^{\pm}(t) = x_{0} \pm v_{p} t$, with velocity $v_{p} = 2p$, where the sign refers to  the left($-$) and right($+$)
 movers. The reason why we have to consider a superposition of two moving packet is intimately connected
 to the geometry of the system; indeed we can think at the initial single particle eigenfunctions like a 
 symmetric superposition of propagative modes $\exp(\pm i p x)$. 
 In our case, the initial particle distribution in the phase-space is
 \be\label{n0_px}
 n_{0}(x,p) = \frac{1}{\pi}\frac{\theta(-x)}{1+\exp(\beta_{\mathcal{L}}p^2)}+\frac{1}{\pi}\frac{\theta(x)}{1+\exp(\beta_{\mathcal{R}}p^2)},
 \ee
 from which we can straightforwardly obtain the time-evolved density profile
 \be
 n(x,t) =\frac{1}{2} \sum_{\sigma=\pm 1} \iint dp\,dx_0 \, n_{0}(x_0,p)\delta(x-x_0-\sigma 2pt).
 \ee
 Injecting Eq.(\ref{n0_px}) in the latter equation, we finally obtain
 \be\label{nxt_scaling}
 n(x,t) = \frac{1}{2}\left\{\begin{array}{cc}  n_{\mathcal{R}} + f_{\mathcal{L}}(x/2t,\infty)+ f_{\mathcal{R}}(0,x/2t) & x>0 \\
 n_{\mathcal{L}} + f_{\mathcal{L}}(0,-x/2t)+ f_{\mathcal{R}}(-x/2t,\infty) & x<0\end{array}\right. ,
 \ee
 where we have introduced the scaling function
 \be
 f_{\mathcal{L/R}}(x,y) \equiv \int_{x}^{y} \frac{dp}{\pi} \frac{1}{1+\exp(\beta_{\mathcal{L/R}} p^2)}. 
 \ee
 
 Following the same lines, we can describe the time-dependent energy-density profile. Indeed, starting from
 the initial energy distribution
  \be\label{e0_px}
 \mathcal{E}_{0}(x,p) = \frac{p^2}{\pi}\frac{\theta(-x)}{1+\exp(\beta_{\mathcal{L}}p^2)}+\frac{p^2}{\pi}\frac{\theta(x)}{1+\exp(\beta_{\mathcal{R}}p^2)},
 \ee
 we straightforwardly obtain
 \be\label{Ext_scaling}
 \mathcal{E}(x,t) = \frac{1}{2}\left\{\begin{array}{cc}  \mathcal{E}_{\mathcal{R}} + g_{\mathcal{L}}(x/2t,\infty)+ g_{\mathcal{R}}(0,x/2t) & x>0 \\
  \mathcal{E}_{\mathcal{L}} + g_{\mathcal{L}}(0,-x/2t)+ g_{\mathcal{R}}(-x/2t,\infty) & x<0\end{array}\right. ,
 \ee
 with 
  \be
 g_{\mathcal{L/R}}(x,y) \equiv \int_{x}^{y} \frac{dp}{\pi} \frac{p^2}{1+\exp(\beta_{\mathcal{L/R}} p^2)}. 
 \ee
 
 In Figure \ref{fig_nE_scaling} we compare the hydrodynamical predictions (\ref{nxt_scaling})
 and (\ref{Ext_scaling}) with the numerically evaluated particle and energy density
 profiles for different system sizes $L$ and rescaled times $t/L$ where one can see a perfect agreement.

 \subsection{Particle current and energy current}\label{sec_currents}
 Using these last results, we can easily derive the particle and the energy currents flowing through the 
 interface separating the two semi-infinite half systems. Indeed, from the continuity equation 
 $\partial_{t} n(x,t) = -\partial_{x} J_{n}(x,t)$ we define the particle current 
 $J_{n}(x,t) \equiv 2\,{\rm Im}[\langle \hat\Psi^{\dag}(x)\partial_x \hat\Psi(x) \rangle_{t}]$. In particular, 
 due to the fact that the current should be vanishing at $x=\pm\infty$, we have
 \be
 J_{n}(x,t) =  -\int_{-\infty}^{x} dz \, \partial_{t} n(z,t).
 \ee
 Therefore, using the scaling form (\ref{nxt_scaling}), the current of
 particle which flows from $\mathcal{L}$ to $\mathcal{R}$ is given by
 \be
 \mathcal{J}_{n} \equiv J_{n}(0,t)  = \frac{\log(2)}{2\pi}\left(\frac{1}{\beta_{\mathcal{L}}} - \frac{1}{\beta_{\mathcal{R}}}\right),
 \ee
 which does not depend on time as expected in the scaling regime. Similarly, the energy current 
 $J_{\mathcal{E}}(x,t) \equiv 2\,{\rm Im}[\langle \partial_x \hat\Psi^{\dag}(x)\partial^2_x \hat\Psi(x) \rangle_{t}]$
 is given by
 \be
 J_{\mathcal{E}}(x,t) = -\int_{-\infty}^{x} dx \, \partial_{t} \mathcal{E}(x,t),
 \ee
 which evaluated at $x=0$, making use of (\ref{Ext_scaling}), gives
 \be
 \mathcal{J}_{\mathcal{E}} \equiv J_{\mathcal{E}}(0,t) = \frac{\pi}{24}\left(\frac{1}{\beta_{\mathcal{L}}^2} - \frac{1}{\beta_{\mathcal{R}}^2}\right).
 \ee
 This result agrees with the Conformal Field Theory (CFT) prediction\cite{dvbd13, bd12} 
 $
 \mathcal{J}^{CFT}_{\mathcal{E}} = [c\pi / (12\hbar v_{F}) ](\beta_{\mathcal{L}}^{-2} 
 - \beta_{\mathcal{R}}^{-2} ),
 $
 with $\hbar =1$, central charge $c=1$ and Fermi velocity $v_{F}=2$. 
Interestingly, in Ref. \onlinecite{kIm13} it has been numerically found that,
 for a large class of problems, the steady-state energy current takes the functional 
 form of a difference between the total radiated power from the left and right leads. 
 However,  new recent results on integrable models of relativistic quantum field theory 
 (IQFT) with diagonal scattering seem to show that this property of the current does not hold in general\cite{cacdh13}.
 
\begin{figure}[t!]
\includegraphics[width=0.5\textwidth]{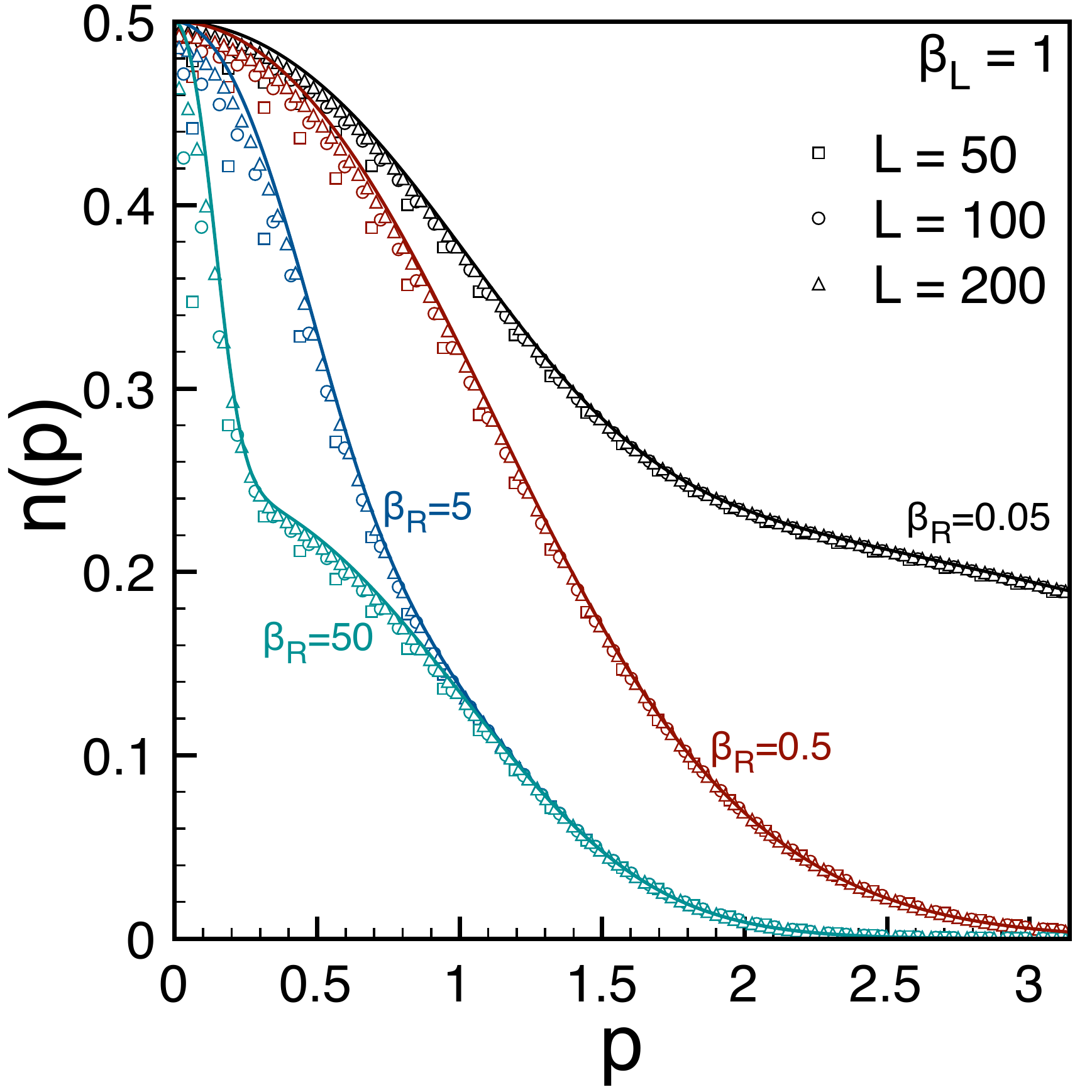}
\caption{Post-quench mode occupation evaluated for different initial temperatures $\beta_{\mathcal{L}}=1$ and 
$\beta_{\mathcal{R}}=0.05,\,0.5,\,5,\,50$. Different symbols represent different sizes $L$ showing the convergence toward Eq.(\ref{n(p)_cont}) (full lines).} 
\label{fig_npGGE}
\end{figure}
\begin{figure*}[t!]
\includegraphics[width=0.33\textwidth]{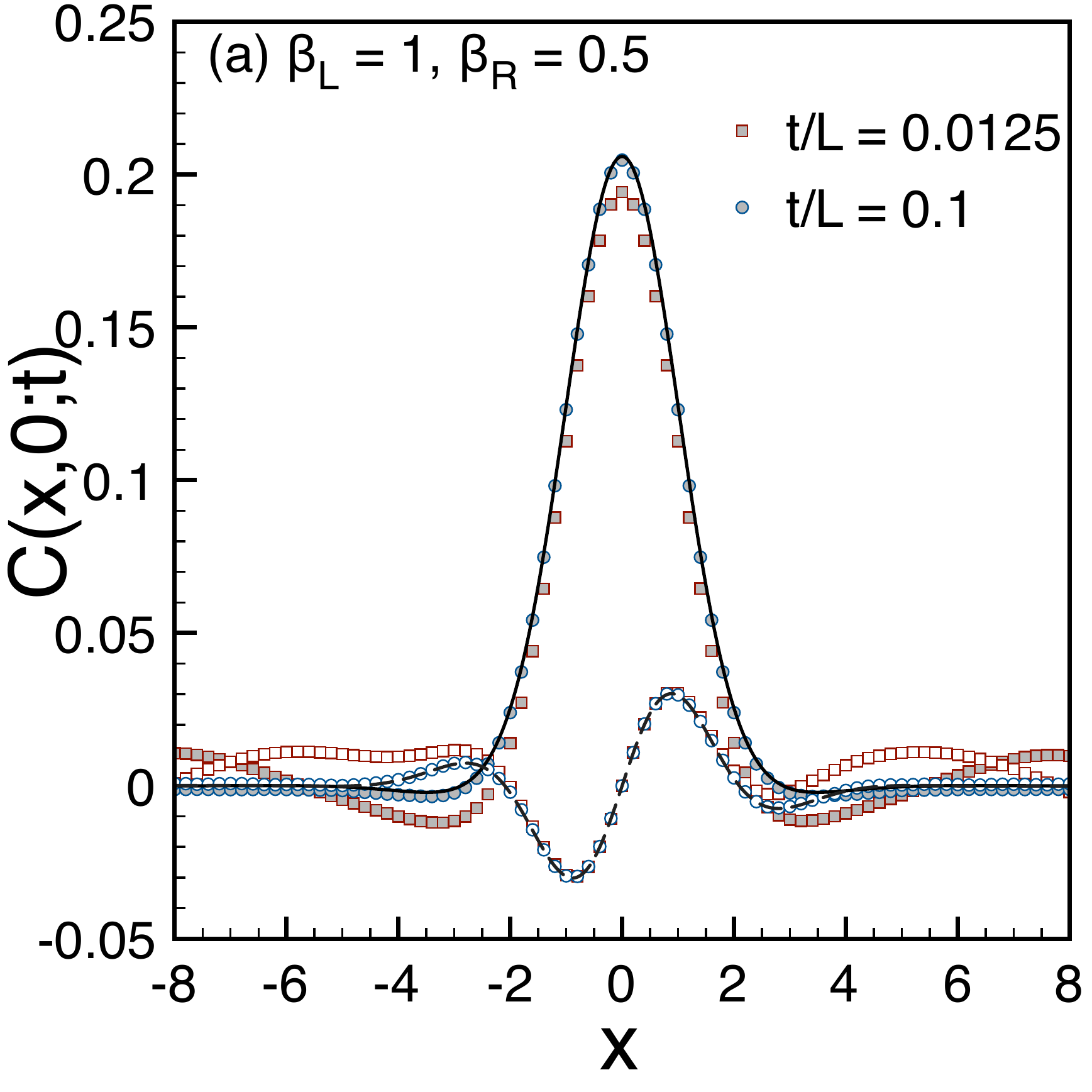}\includegraphics[width=0.33\textwidth]{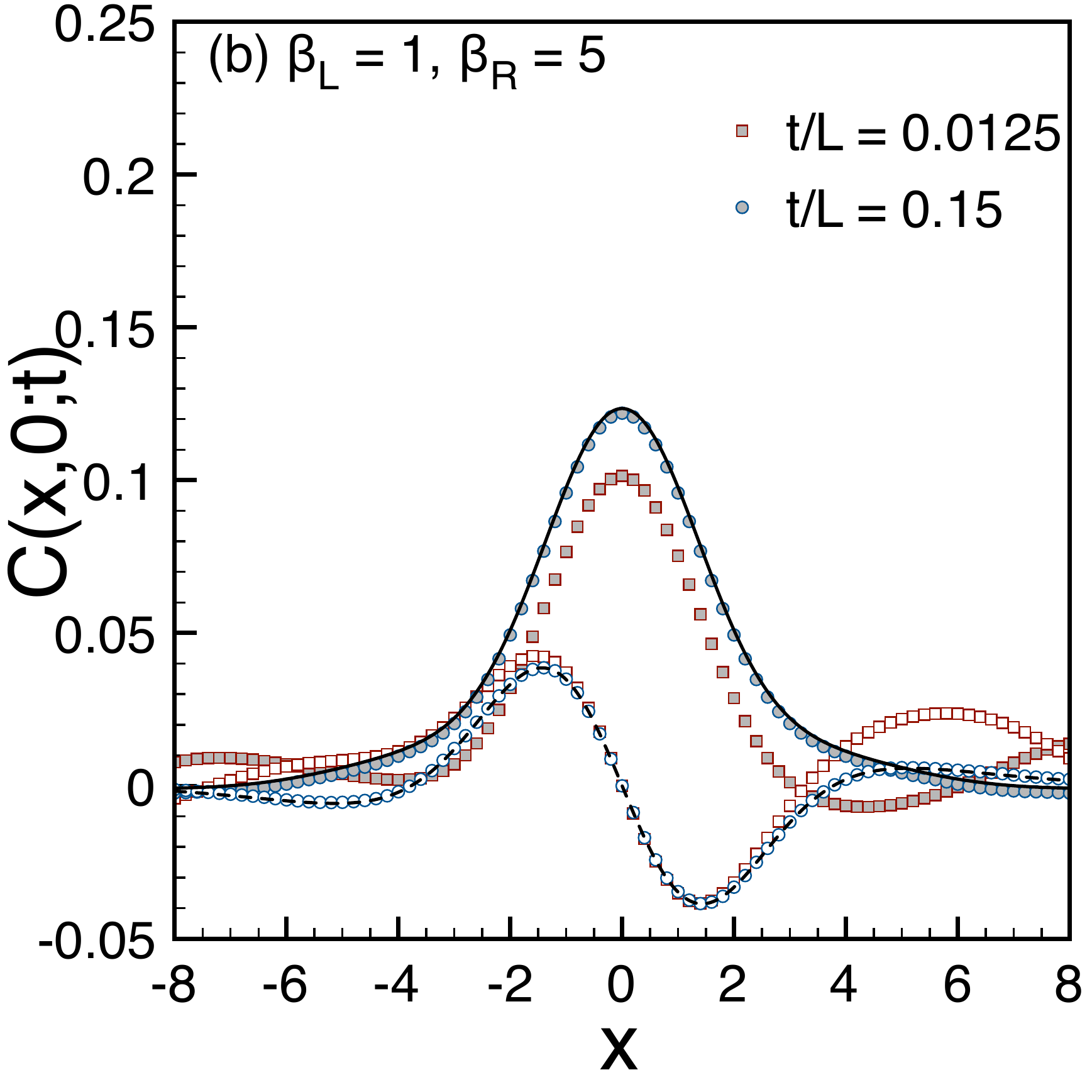}\includegraphics[width=0.33\textwidth]{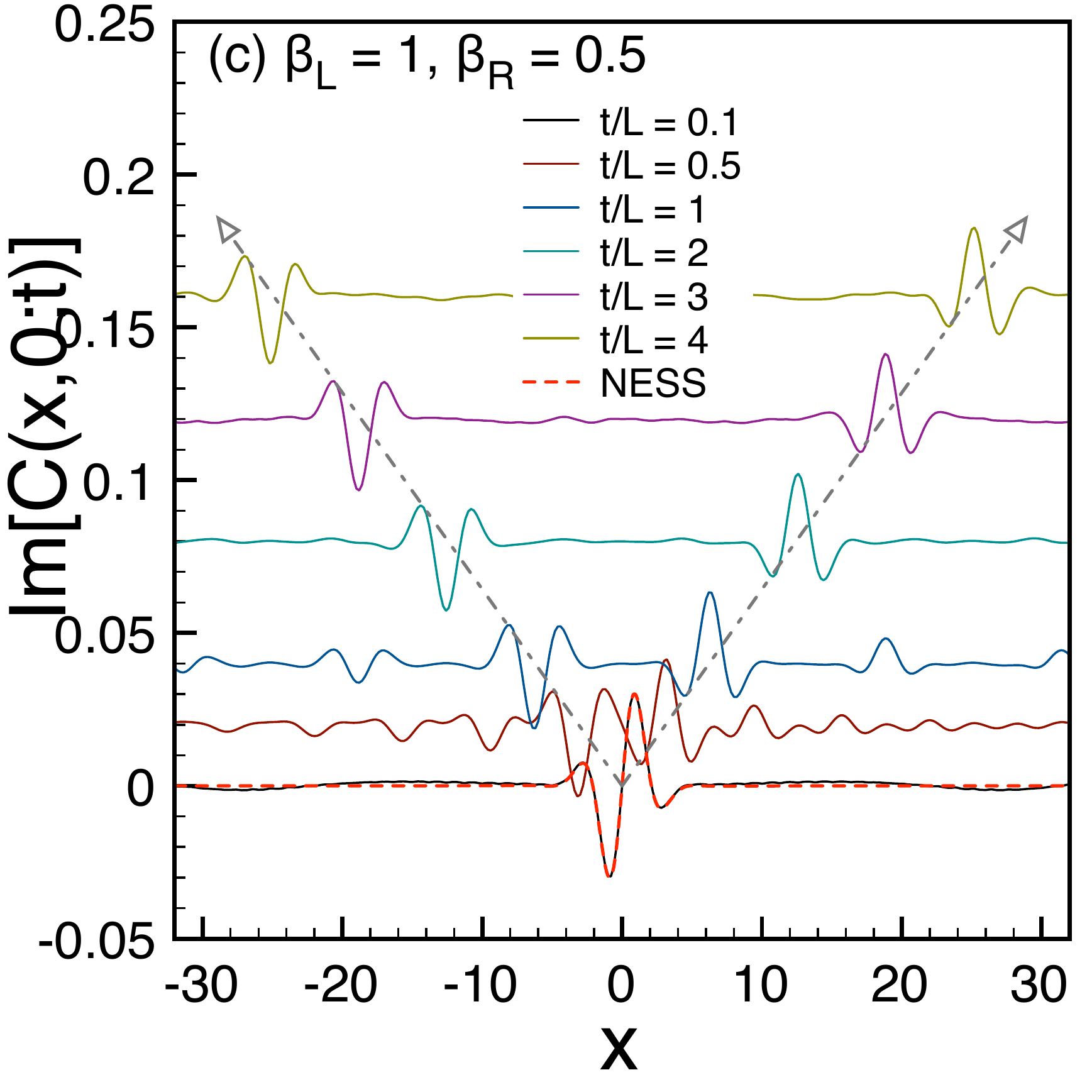}
\caption{Numerically evaluated correlation function $C(x,0;t)$ for a system with size $L=400$ initially 
prepared with $\beta_{\mathcal{L}}=1$ and $\beta_{\mathcal{R}}=0.5$ (a) and $\beta_{\mathcal{R}}=5$ (b). 
Different symbols correspond to different times (filled symbols for the real part, empty symbols 
for the imaginary part). The lines are the real (full lines) and imaginary (dashed lines) part  
of the analytic correlation function in the non-equilibrium steady-sate given by Eq. (\ref{Corr_stat}). 
(c) The behavior of the imaginary part of the correlation function evaluated 
for different rescaled times $t/L$.
As expected, if $t\ll L$, the numerically evaluated correlation function agrees with the 
NESS prediction in (\ref{Corr_Im}) . Otherwise, for larger rescaled time $t/L\gg 1$ the correlation function takes the shape of two traveling peaks with velocity $w=2\pi/L$. For the sake of clarity the curves are vertically 
shifted by an amount $\delta_{shift}= 0.04 (t/L)$.   
Notice the perfect agreement with peak's law of motion $x_{peak}(t/L) = 2\pi (t/L)$.}
\label{fig_corrxt}
\end{figure*}
 
 \subsection{Large-time limit and boundary effects}
\subsubsection*{Non-Equilibrium Steady State (NESS)}
The description of the non-equilibrium stationary state in an infinite quantum system was done for the $XY$ chain in Ref. \onlinecite{araki, ogata, aschbacher}  with full mathematical rigor in the framework of $C^*$-algebraic dynamical systems. 
However, in our case,  to avoid boundary effects and give a long-time description of the local physics around the connecting point we proceed in the following way: $(i)$ we 
rewrite the two-point correlation function in the TD limit ($L\rightarrow \infty$), changing all sums to integrals; 
$(ii)$ and then take  $t\to\infty$ limit which allows us to use the stationary-phase approximation\cite{olver,wong}. 
Following this recipe, the two-point correlation function in the stationary regime is found to be given by
\be\label{Corr_stat}
C^{Stat}(x,y) = C^{Stat}_{Re}(x-y) + i \, C^{Stat}_{Im}(x-y)\; 
\ee
where the real part 
\be\label{Corr_Re}
C^{Stat}_{Re}(z) = \int_{0}^{\infty} \frac{dp}{\pi} \cos(pz) \, n(p),
\ee
with
\be\label{n(p)_cont}
n(p) = \frac{1}{2}\left[ \frac{1}{1+\exp(\beta_{\mathcal{L}} p^{2} )} 
 +\frac{1}{1+\exp(\beta_{\mathcal{R}} p^{2} )}  \right],
\ee
and 
\be\label{Corr_Im}
C^{Stat}_{Im}(z) = \int_{0}^{\infty} \frac{dp}{2\pi} 
\left[ \frac{\sin(pz)}{1+\exp( \beta_{\mathcal{R}} p^2 )} - \frac{\sin(pz)}{1+\exp( \beta_{\mathcal{L}} p^2 )} \right]\; ,
\ee
for the imaginary part. 
The  $n(p)$ is the TD limit of the initial mode occupation $\langle \hat n _{p_n}\rangle_0$, 
i. e. given by the diagonal terms of Eq. (\ref{etaeta_0}).  In Figure \ref{fig_npGGE}, 
we plot $\langle \hat n_{p_n} \rangle_{0}$ for different system sizes, showing the convergence toward the 
TD-limit result.
Introducing the two-sided Fermi distribution 
$\tilde f(p)\equiv \theta(p) \tilde f_{\mathcal{L}}(p) +\theta(-p) \tilde f_{\mathcal{R}}(p)$, 
where the momentum is now considered on the whole real axis and with
$\tilde f_{\mathcal{L/R}}(p)=\frac{1}{1+ e^{\beta_{\mathcal{L/R}}p^2}}$, one can also rewrite the stationary correlation function as the Fourier transform of $\tilde f(p)$:
\be
C^{Stat}(x,y)=\int_{-\infty}^\infty  \frac{dp}{2\pi} \tilde f(p)\,  e^{-ip(x-y)} 
\ee
such that 
\be
C^{Stat}_{Re}(z) = \int_{-\infty}^\infty  \frac{dp}{2\pi} \tilde f(p) \, \cos (pz)
\ee
and 
\be
C^{Stat}_{Im}(z) = - \int_{-\infty}^\infty  \frac{dp}{2\pi} \tilde f(p) \, \sin (pz)\, .
\ee
As a consequence of this redefinition of the momenta $p$, the non-equilibrium stationary state (NESS) described by this correlation function can be viewed as a superposition of left and right stationary movers, each of them being distributed with respect to there own inverse temperature $\beta_{\mathcal{L}}$ and $\beta_{\mathcal{L}}$. 
In this regime (large time limit and close to the connecting point) translation invariance is restored and the stationary density is uniform over the whole region with $\overline{n} = (n_{\mathcal{L}}+n_{\mathcal{R}})/2$. The imaginary part gives rise to non-vanishing stationary currents 
 (reproducing exactly the results already found in the previous section for $\mathcal{J}_{n}$ and  
 $\mathcal{J}_{\mathcal{E}}$) which highlight the nature of the NESS
 describing the system. 
 
 Once again, we want to stress that 
 Eq. (\ref{Corr_stat}) which represents the
 correlation function in the NESS, was found considering an infinitely extended
 system, i. e. $L\to\infty$, from the very beginning of the calculation. But of course, the numerical evaluation of the correlation function is done using finite system sizes and as a consequence, the NESS description applies only if $t\ll L$.
 In Figure \ref{fig_corrxt} we show for two different initial conditions
 how the time-dependent correlation function approaches the stationary value in a region around the origin.
 Notice that we can not consider large value of $t/L$ if we want to be coherent with the above description of the 
 steady-state.
\begin{figure}[t!]
\includegraphics[width=0.25\textwidth]{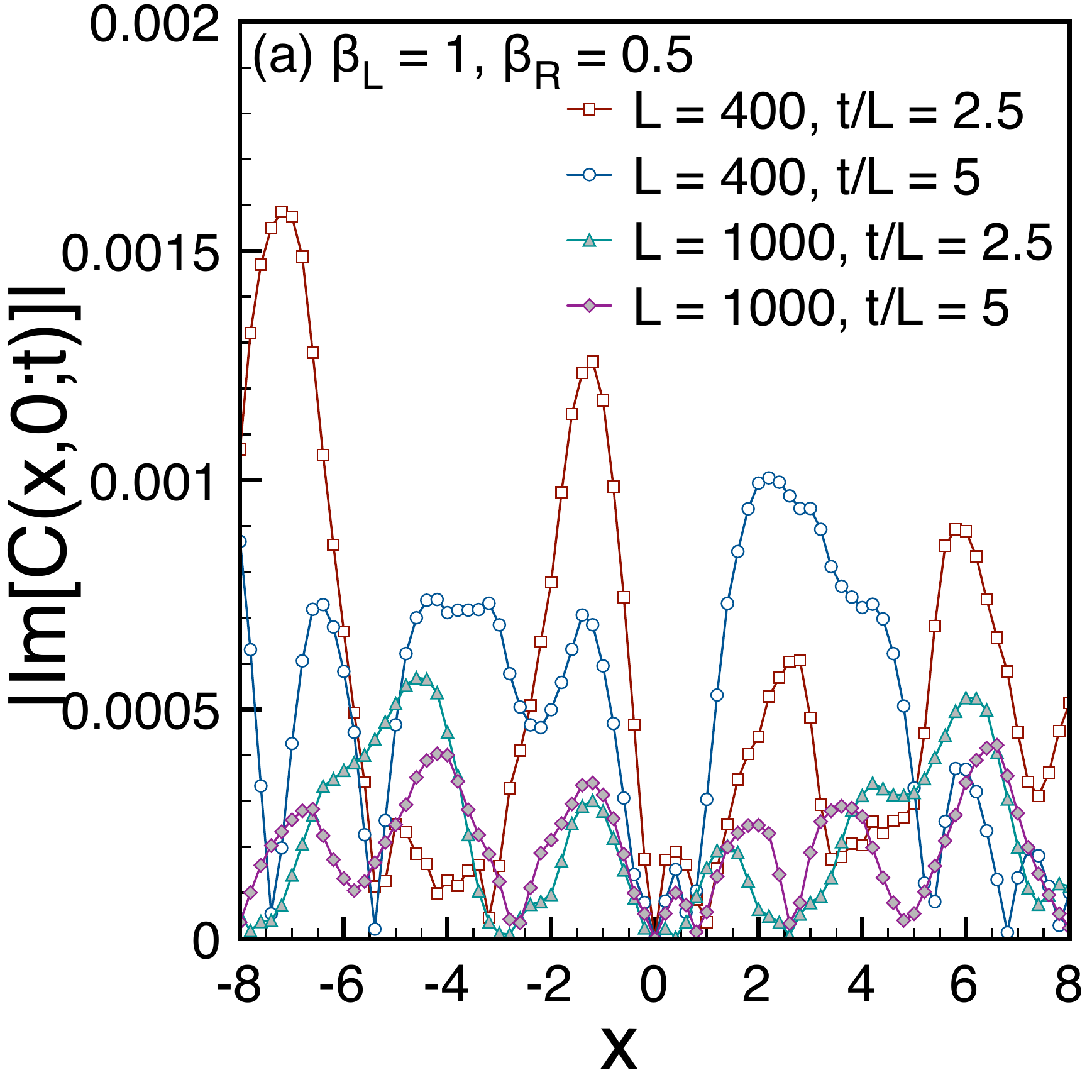}\includegraphics[width=0.25\textwidth]{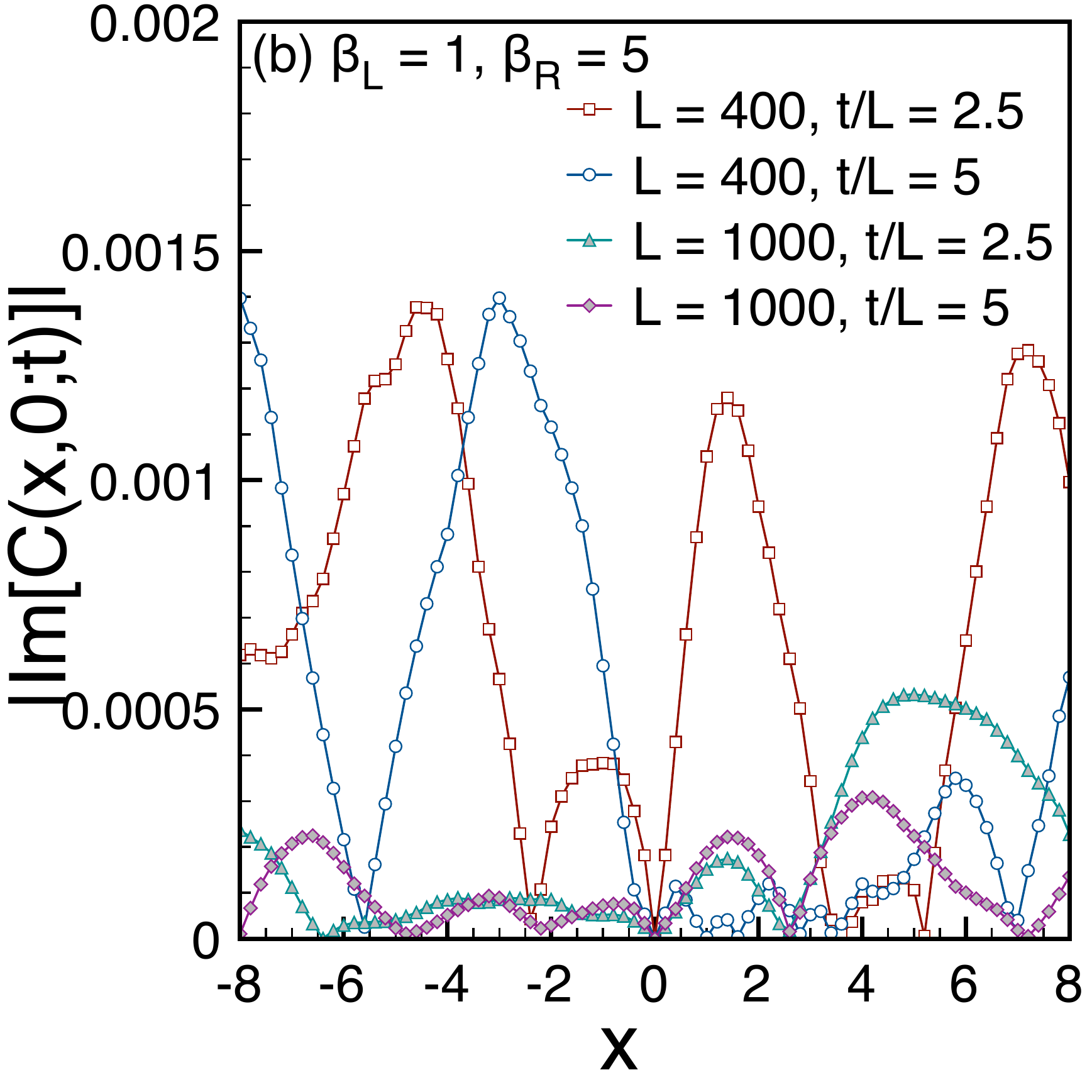}
\includegraphics[width=0.25\textwidth]{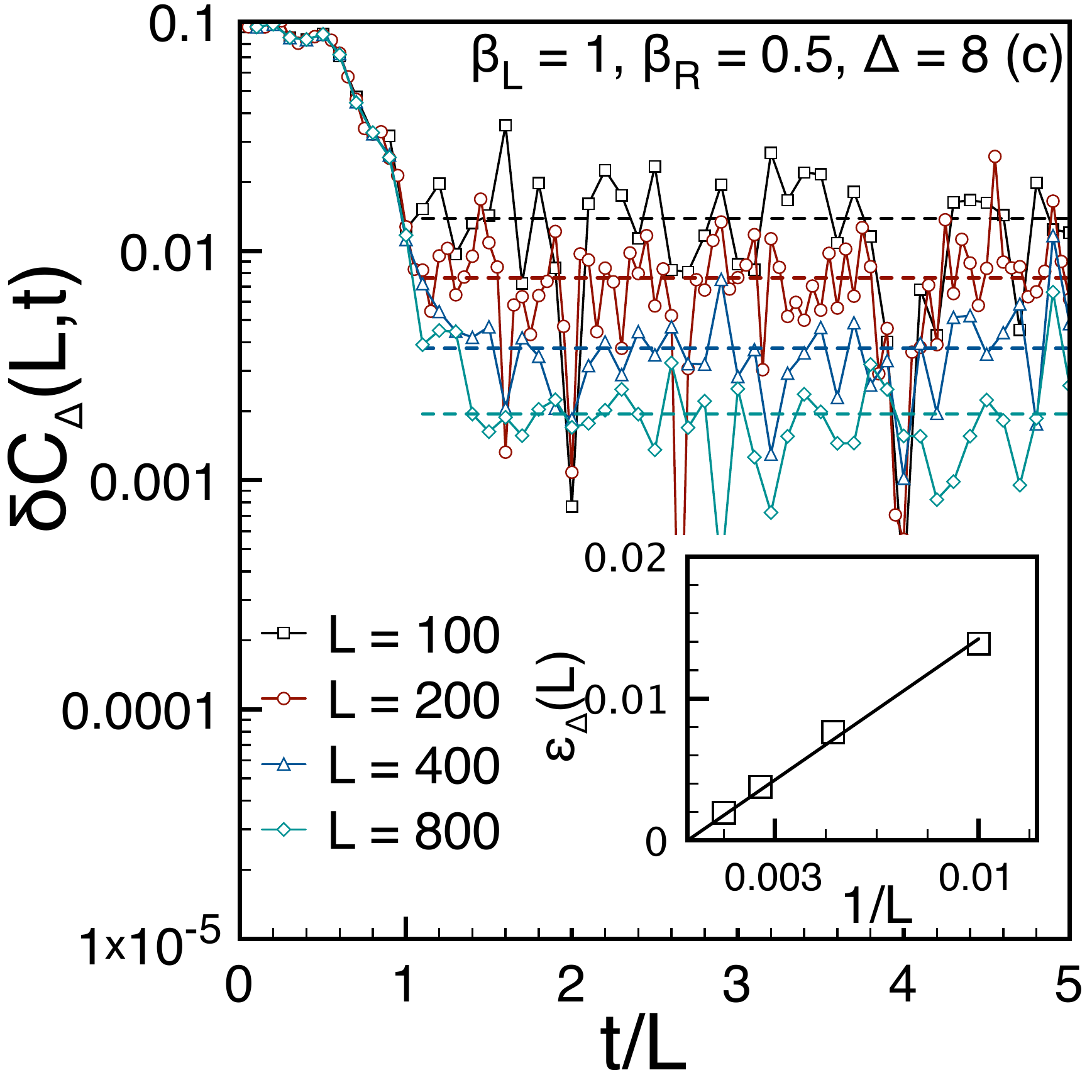}\includegraphics[width=0.25\textwidth]{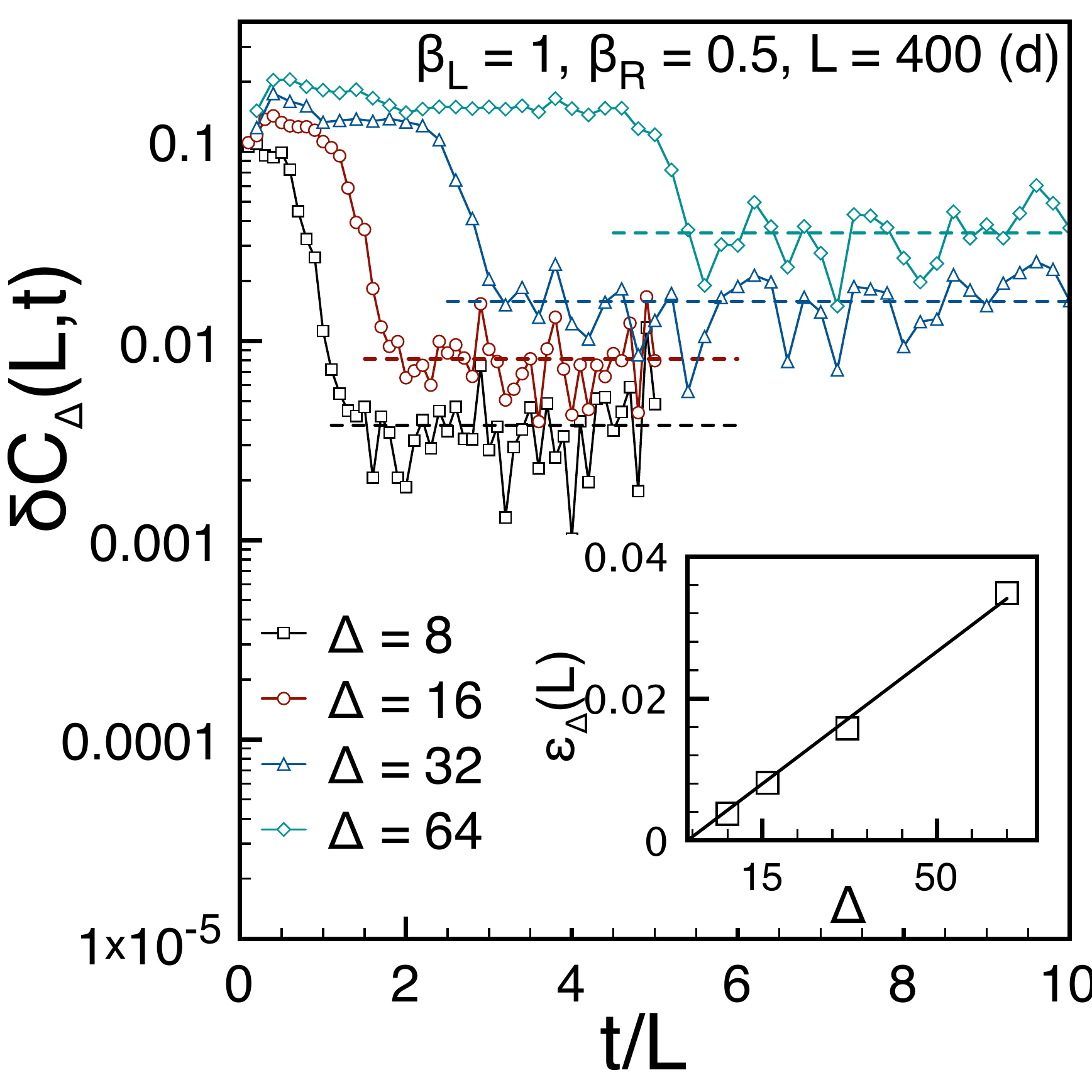}
\caption{(a,b) The absolute value of the imaginary part fo the correlation function numerically evaluated 
for different $L$ and $t/L$. At fixed $t/L\gg 1$, we expect that ${\rm Im}C(x,y;t) \to 0$ for $L\to\infty$.  
As a support of this, notice how for the times reported in figure, the fluctuations are $\lesssim 0.0015$ for
$L=400$ and for $L=1000$ they become $\lesssim 0.0005$ giving a numerical evidence that 
${\rm Im}[C(x,y;t)]$ is approaching zero. (c,d) Integrated imaginary fluctuations $\delta C_{\Delta}(L,t)$ as
a function of the rescaled time $t/L$ for: different sizes $L$ and fixed $\Delta=8$ (c); 
fixed $L=400$ and different subdomains $\Delta$ (d). 
The dashed lines represent the qusi-bounds $\varepsilon_{\Delta}(L)$ defined in the main text.
Finally, in the insets of (c,d) we show the scaling of $\varepsilon_{\Delta}(L)$ respectively vs $1/L$ 
or vs $\Delta$.}
\label{fig_ImCorr}
\end{figure}

\subsubsection*{Equilibration regime}
If one extends the numerical evaluation of the post-quench 
 correlation function for times such that $t > L$ then a completely new phenomenon 
 appears (see Figure \ref{fig_corrxt} (c)): the imaginary part of the correlation function takes the shape 
 of two traveling peaks which move with velocity $w=2\pi/L$. 
 The proper rescaled time in order
 to take trace of such a behavior is $t/L$: in terms of this ``new time'',  the peaks propagate in space
 with a ``rescaled velocity'' $\tilde w = 2\pi$ which remains finite.
 In the region between these two moving peaks the imaginary part of the correlation function vanishes
 a part from small fluctuations due to finite size effects. This behavior is reminiscent of what was already
 observed in Ref. \onlinecite{csc13} and shows the crucial role of the boundaries for the system equilibration.  In the following we analyze in details this equilibration regime.
 
  At large times, $t\gg L$ (with $t/L^2 \ll 1$ to avoid revivals), the free boundaries of the system matter for the properties of the correlation function $C(x,y)$ in the vicinity of the connection point. In this limit, 
in the double sum of Eq.(\ref{Corr_t}) only the diagonal terms
 $m=n$  survive \footnote{Notice that the terms $m=-n$ are not allowed since $m$ and 
 $n$ should be both nonnegative.}. 
 Therefore, the large-time limit of the correlation function in a system 
 with boundaries reduces to just the real contribution
 \footnote{Notice that this result exactly coincides with the time average 
 $\lim_{T\to\infty} T^{-1}\int_{0}^{T} dt \, C(x,y;t)$}.
\be\label{Corr_avg}
C_{\infty}(x,y) = \sum_{n=0}^{\infty} \varphi_{p_n}(x)\varphi_{p_n}(y) \langle \hat n_{p_n} \rangle_{0} 
\equiv C^{Stat}_{Re}(x-y)\; .
\ee
In other words, the role of the boundaries is to elastically reflect the particles. Moreover, particles with different momenta will reach the edges at different times causing a  global dephasing which increases with time \cite{bs08,csc13}. 
In this regime, at large rescaled times $t/L$, any current will be suppressed which is reflected in the vanishing of the imaginary part of the correlation function as seen in Figure \ref{fig_corrxt} (c).
We further support this picture by means of a numerical analysis showed in Figure \ref{fig_ImCorr} (a,b)
 where we plot the absolute value of the imaginary part of the correlation function for large rescaled times and
 two different initial conditions, i.e. $\beta_{\mathcal{L}}=1$, $\beta_{\mathcal{R}}=0.5$ and  
 $\beta_{\mathcal{L}}=1$, $\beta_{\mathcal{R}}=5$. It is evident from this figure that, for sufficiently large 
 rescaled times $t/L$, the imaginary part of the two-point function vanishes a part from small fluctuations 
 which are suppressed as the system size $L$ increases. 
 
 In order to make this qualitative behavior more
 quantitate, we introduce a measure of these fluctuations by integrating the absolute value of 
 $ {\rm Im}[C(x,0;t)$] in a local domain $[-\Delta/2,\Delta/2] \subset [-L/2,L/2]$,
 \be
 \delta C_{\Delta}(L,t) \equiv \int_{-\Delta/2}^{\Delta/2} dx \, \left| {\rm Im}[C(x,0;t)] \right|.
 \ee
Numerical evidences reported in Figure \ref{fig_ImCorr} (c,d) show that for $1\ll t/L\ll L$ the absolute deviation 
 $\delta C_{\Delta}(L,t) \lesssim \varepsilon_\Delta(L)$ is bounded 
where the bound $\varepsilon_\Delta(L)$ shows a numerical scaling $\sim \Delta/L$. As a consequence,
if the local domain $\Delta$ is scaled proportionally to the system size, the perfect equilibration toward
 Eq. (\ref{Corr_avg}) will be never seen. 
 This shows that equilibrium, where the correlators are given by (\ref{Corr_avg}),  is reached only in a local sense, that is when the size $\Delta$ of the domain considered around the origin is such that $\Delta(L) \sim o(L)$.
  
Finally, another interesting feature evident from Figure \ref{fig_ImCorr} (d) is that the time $t$ required 
 for the correlation function to reach a stationary value in the local domain $[-\Delta/2,\Delta/2]$ increases with $\Delta$,
 as it should be. Indeed, remember that the peaks in ${\rm Im}[C(x,y;t)]$ move with a velocity $w=2\pi /L$. 
 For that reason, the larger is the local domain wherein one analyzes the correlation function, 
 the larger is the time one has to wait in order to expel out the peaks and see the stationary behavior.

Summarizing, once $\Delta$ is fixed, we can individuate four regimes in the time-evolution:
\begin{itemize}
\item[\it i)] {\it an initial transient} during which the correlation function is not yet translational invariant and 
 it is approaching the non-equilibrium steady-state; 
\item[\it ii)]  {\it a  NESS regime}, for $t\gg 1$ but $t/L \ll 1$, corresponding locally to the non-equilibrium 
 current-carrying state, wherein the correlators are given by Eq. (\ref{Corr_stat});
\item[\it iii)]  {\it an intermediate regime} whose duration extends up to the rescaled time 
 $(t/L)^{*} \simeq \Delta / (2\tilde w)$ and which correspond to the 
 intermediate almost constant plateaus in Figure \ref{fig_ImCorr} (d): in this regime the non-equilibrium 
 correlators (namely Eq.(\ref{Corr_Im})) are already destroyed but the moving 
 peaks are still inside the region $[-\Delta/2,\Delta/2]$;
 \item[\it iv)] {\it finally, the equilibrium stationary state} for $t\gg L$, with no more currents and characterized by the correlation function in Eq. (\ref{Corr_avg}). 
\end{itemize}
 Notice that, the first two regimes refer to the behavior of the correlation function
 reported in Fig. \ref{fig_ImCorr} (c,d) for $t/L \sim 0$ and, therefore, are suppressed in that figure.
 Once again, let us emphasize that the regime iv) will be never reached if $\Delta \sim L$ or if we remove the boundary effects
 from the very beginning of the calculation. Na\"ively, the last condition simply means that 
 if we consider $L\to\infty$ from the very beginning then $t/L \ll 1$ for any finite $t$ and only 
 the regimes i-ii) will survive.
 
\section{The reduced density matrix and Statistical Ensembles}\label{GGE}
\subsection{Generalized Gibbs Ensemble}
In the previous section we have found the analytical form of the stationary two-point correlation function 
in the large-time limit. We have seen that, whenever a proper time-space scaling is done, in order to
retain the effects of the boundaries, this function depends only on the post-quench mode occupation 
$n(p)$, and thanks to the Wick's theorem, all other observables can be described in terms of $n(p)$. 
With some abuse of language, we can say that the stationary state of the system after the quench can be
described by the so called Generalized Gibbs Ensemble (GGE) \cite{rdyo07,rdo08,r09,rs12}
 \be\label{GGE_local}
\hat\rho_{GGE} = Z_{GGE}^{-1}\exp\left\{ - \sum_{j} \gamma_j \hat I_{j}\right\} 
\equiv Z_{GGE}^{-1} \, {\rm e}^{-\hat H_{eff}},
 \ee
 where $\hat I_{j} $ are  local integral of motion, $\rm{Tr}\hat\rho_{GGE} =1$, and where we defined 
 an ``effective'' Hamiltonian $\hat H_{eff}$. The Lagrange
 multipliers $\gamma_{j}$ are fixed by the initial state trough the conditions 
 $\rm{Tr}[\hat I_j \hat\rho_{GGE}]=\langle \hat I_j \rangle_0$. However, whenever a closed system evolves
 unitarily, even if the initial state is not  prepared in a pure state, to talk about a stationary state 
 $\hat\rho_{GGE}$ describing the whole system may seem paradoxical. This paradox is resolved using the
 reduced density matrix $\hat\rho_{A}(t)\equiv \rm{Tr}_{B}[\hat\rho(t)]$, where $B$ is the complement of $A$
 and $\hat\rho(t) = \exp(-i \hat H t) \hat\rho_{0} \exp(i \hat H t) $ is the time-evolved density matrix of the whole
 system \cite{cdeo08,bs08,CEF,scc09,r09,rsm10}. 
 Indeed, we should think at the GGE in a ``local'' sense saying that the system reaches a stationary state
 if, after properly taking the TD limit, the limit $\hat\rho_{A,\infty} \equiv \lim_{t\to\infty}\hat\rho_{A}(t)$
 exists for any finite $A$. Then, we say that it is described by a statistical ensemble $\hat\rho_{E}$ if
 the reduced density matrix $\hat\rho_{A,E} \equiv \rm{Tr}_{B}[\hat\rho_{E}]$ equals $\hat\rho_{A,\infty}$. 
 In practice, carrying out the time evolution of the reduced density matrix is not a trivial task. 
However, for a fermionic quadratic theory, it has been shown that the reduced density matrix can be
written as an exponential of the fermionic fields\cite{cmv11,peschel,pe09}
\be
\hat\rho_{A} = Z_{A}^{-1} \exp \left\{ -\int_{A} dx dy \, \hat\Psi^{\dag}(x) S_{A}(x,y;t) \hat\Psi(y) \right\},
\ee
where the function $S_{A}(x,y;t)$ is connected to the restriction over the region $A$ of the correlation matrix 
$C(x,y;t)\equiv \langle \hat\Psi^{\dag}(x) \hat\Psi(y)\rangle_{t}$ via the formal relation
\begin{equation}\label{S_to_C}
S_{A}(x,y;t) = \ln \frac{\delta_{A}(x-y) - C_{A}(x,y;t) } { C_{A}(x,y;t)},
\end{equation}
where $C_{A}(x,y;t) = C(x,y;t),\,\forall x,y\in A$,
and $\delta_{A}(x-y)$ is the Dirac Delta function restricted in $A$. Thanks to this fact, if the two-point 
correlation function at large times is described by a statistical ensemble, then the expectation value of any
local observable will also be.
 
\begin{figure}[t!]
\includegraphics[width=0.5\textwidth]{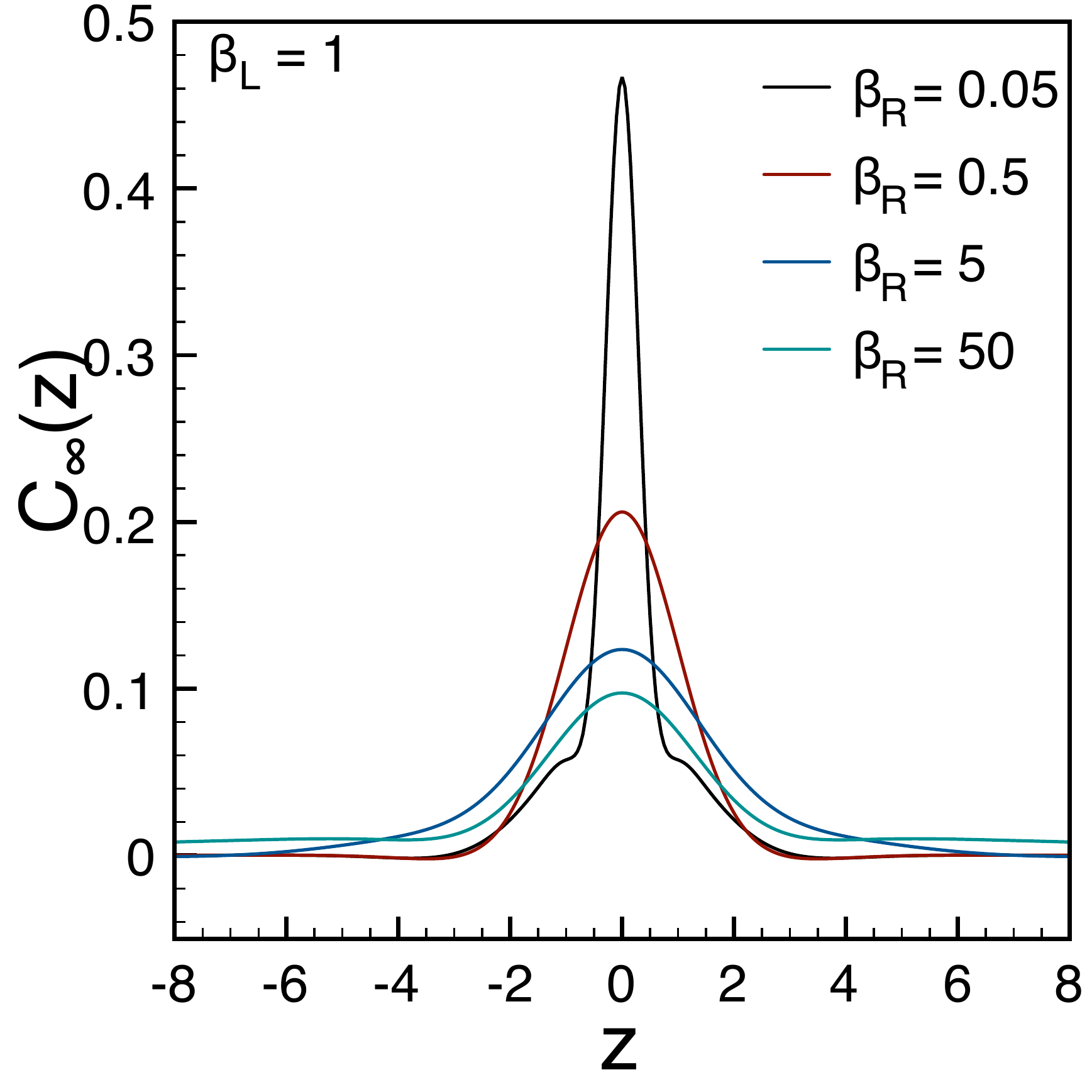}
\caption{Stationary correlation function $C_{\infty}(z)$, which agrees with the GGE correlation function,
 for a system with boundaries initially prepared with $\beta_{\mathcal{L}}=1$ and 
$\beta_{\mathcal{R}}=0.05,\,0.5,\,5,\,50$.} 
\label{fig_corrGGE}
\end{figure}

For free fermion models, instead of using the local charges $\hat I_{j}$, we can work with the 
post-quench occupation modes $\hat n_{p_{n}}$, which are linearly connected with the local charges \cite{ck12, fe13,csc13}. 
Therefore, in the quench from a tensor thermal state in a non-interacting fermionic theory, the GGE will be
\bea\label{GGE_mode}
\hat\rho_{GGE} & = & Z_{GGE}^{-1} \exp \left\{ - \sum_{n=0}^{\infty} \lambda_{p_n} \hat n_{p_n}\right\},\\
 & = &  Z_{GGE}^{-1} \exp \left\{ - L \int_{0}^{\infty} \frac{dp}{\pi}\, \lambda(p) \hat n(p)\right\},
\eea
where the lagrange multipliers are determined via $[1+\mathrm{e}^{\lambda(p)}]^{-1} = n(p)$.
In Figure \ref{fig_corrGGE} we plot the GGE correlation function 
$C_{GGE}(x,y)\equiv {\rm Tr} [\hat\Psi^{\dag}(x)\hat\Psi(y) \hat\rho_{GGE}] = C_{\infty}(x,y)$, for
some initial conditions. 

It is worth mentioning that, expanding $\lambda(p)$ around zero and using the parity of $n(p)$,
 one can rewrite the Eq. (\ref{GGE_mode}) in terms of the local charges 
\bea
\hat I_{2j} & \equiv & (L/\pi) \int_{0}^{\infty} dp\,p^{2j} \hat n(p) \\
& = & \iint dx dy \, (-1)^{j} \hat\Psi^{\dag}(x) \delta^{(2j)}(x-y) \hat\Psi(y) \\
& = & \int dx \, \hat\Psi^{\dag}(x) (-1)^{j} \partial^{2j}_{x} \hat\Psi(x),
\eea
with Lagrange multipliers $\gamma_{2j} = (\partial_{p}^{2j} \lambda(p)|_{0})/(2j)!$, 
obtaining for the ``effective'' Hamiltonian the following expression 
\bea\label{H_eff}
\hat H_{eff} & = & \sum_{j} \gamma_{2j} \hat I_{2j}\\
& = & \int dx\, \hat\Psi^{\dag}(x) \bigg[ \sum_{j} (-1)^{j} \gamma_{2j} \, \partial^{2j}_{x}\bigg] \hat\Psi(x).
\eea
Now some comments are due. As it should, $\hat H_{eff}$ is local in the sense that 
it is an integral of a local current. Nevertheless, the differential operator 
$\sum_{j} (-1)^{j} \gamma_{2j} \, \partial^{2j}_{x}$
acting on the local fields contains all the even derivatives: in other words, depending on 
the behavior of $\gamma_{2j}$, the ``effective'' interaction could be long-range.

It is straightforward to show
 that $\gamma_{0}=0$ and $\gamma_{2}=(\beta_{\mathcal{L}}+\beta_{\mathcal{R}})/2$. This result states
that, in the asymptotic stationary state, the Lagrange multiplier associated to the post-quench Hamiltonian 
($\hat H \equiv \hat I_2$), i.e. the ``equilibrium inverse temperature'',
equals the average of the initial inverse temperatures. This particular result is an effect of the coexistence 
of the entire set of conserved charges and, as we will see in the next paragraph, it will not survive if one
considers a truncated set of local charges. Moreover, it can be easily shown that $\gamma_{2j}$ decays 
exponentially for large $j$ making the ``effective'' Hamiltonian $\hat H_{eff}$ short-range interacting, 
in contrast to what was observed in Ref. \onlinecite{ogata} for the non-equilibrium stationary state 
following a similar quench in the XX model, where it was found that the
effective Hamiltonian is long-range interacting.
\footnote{Indeed, in terms of the local conserved charges of the XX-model, i.e.  
$\hat Q^{\pm}_{j} = i^{(1\mp1)/2}\sum_{i} (\hat c^{\dag}_{i} \hat c_{i+j} \pm h.c.)/2$, 
the effective Hamiltonian takes the form 
$\hat H_{eff} = \hat H_{XX} + \sum_{j} \mu^{-}_{j} \hat Q^{-}_{j}$, with $\mu^{-}_{j}\sim 1/j$
for large $j$, where  $\hat c_{j}\,(\hat c^{\dag}_{j}$) are the Fermi annihilation (creation)
operators on lattice and $\hat H_{XX}$ is the XX-model hamiltonian. 
However, we remand to Ref. \onlinecite{ogata} for further details.}

Once again, we want to stress that the description of the large-time stationary-state 
via the GGE is correct only if the effect of the boundaries is retained in the TD limit.
 In other words, the proper scaling which permits to see a true equilibrium state (without any current) is 
 obtained only by considering $L\to\infty$ with $t/L\gg 1$ and $t/L^2 \ll 1$.
\begin{figure}[t!]
\includegraphics[width=0.25\textwidth]{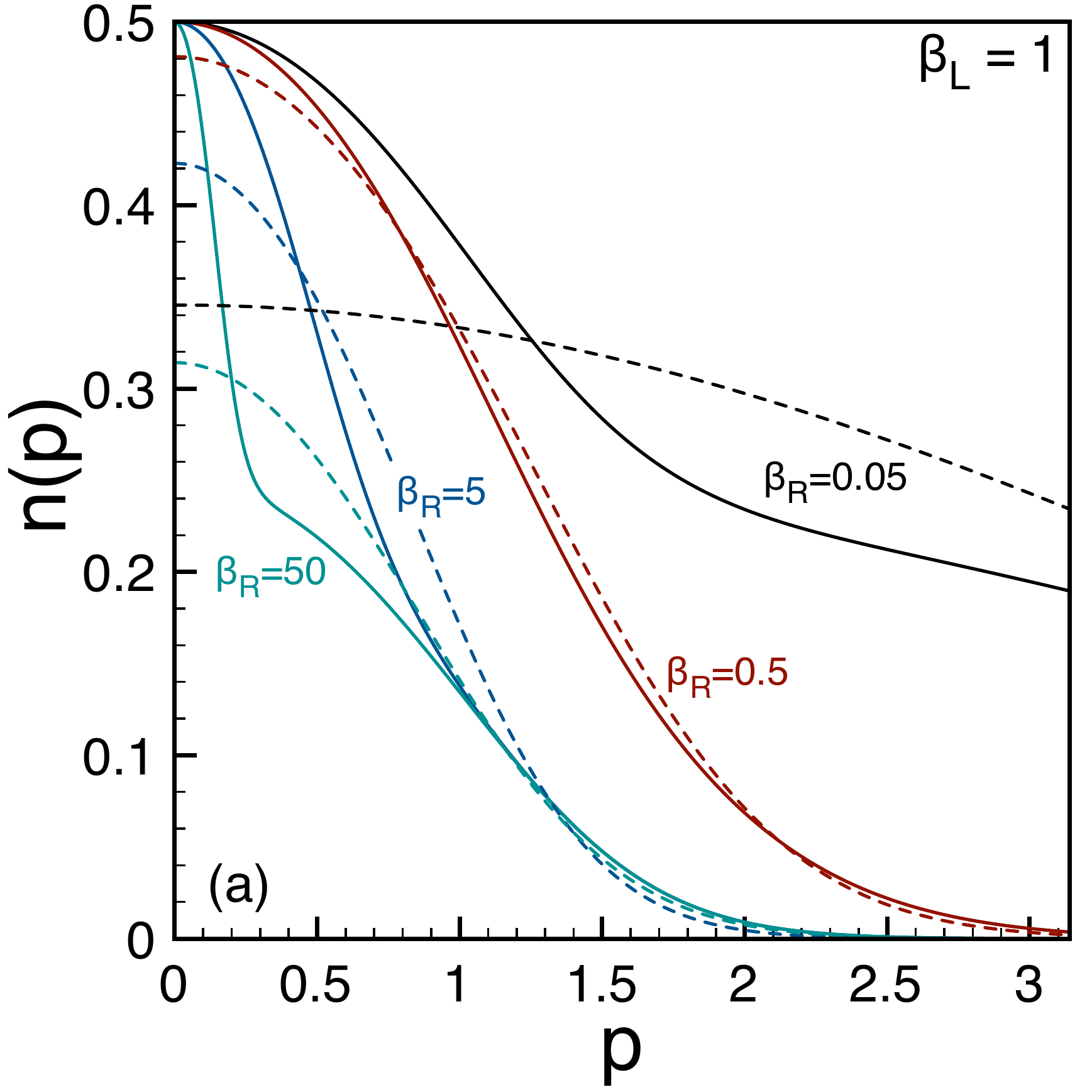}\includegraphics[width=0.25\textwidth]{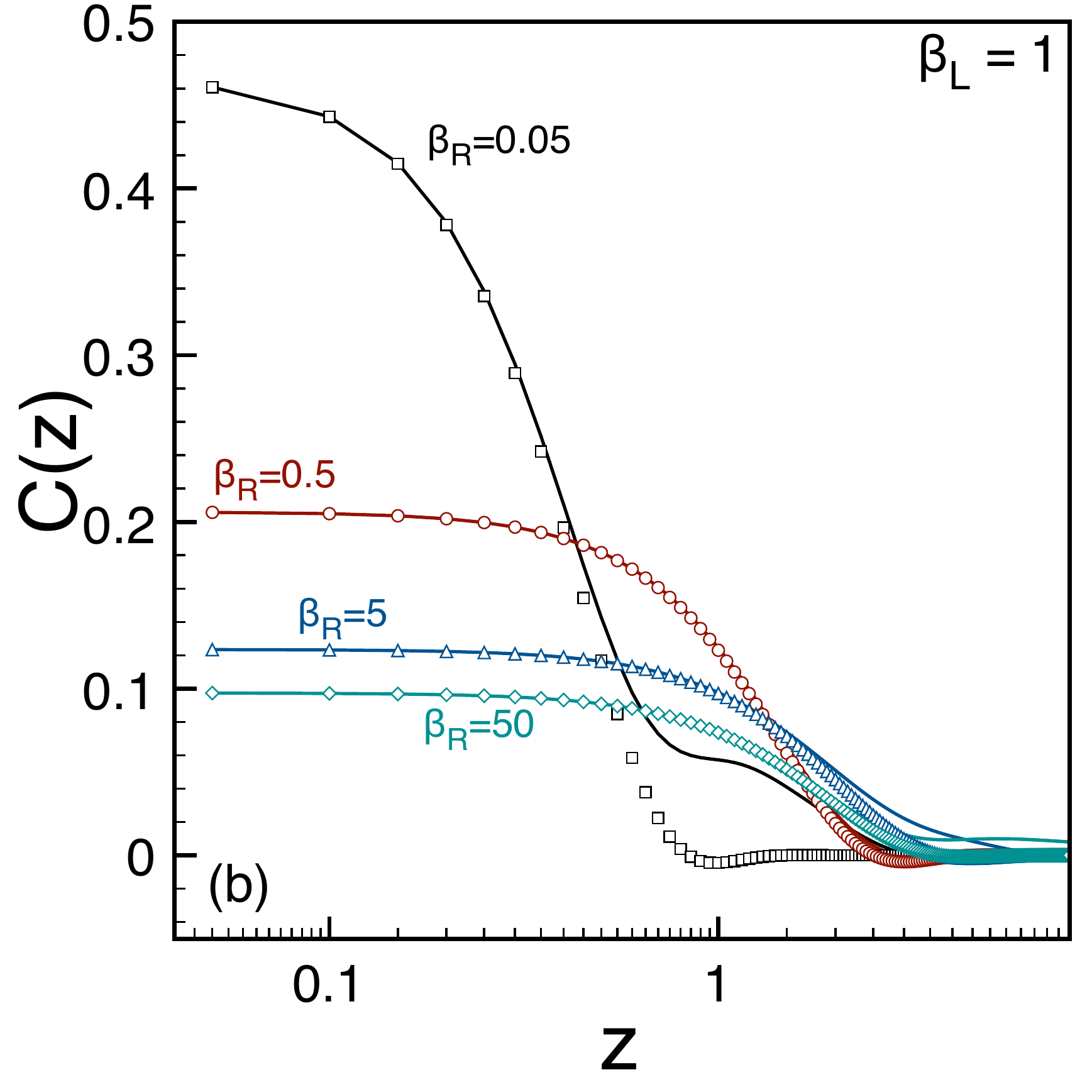}
\caption{(a) Post-quench mode occupation in the GGE (full lines) and in the GCE (dashed lines)
evaluated for different initial temperatures $\beta_{\mathcal{L}}=1$ and $\beta_{\mathcal{R}}=0.05,\,0.5,\,5,\,50$. 
(b) Two-point correlation function in the GGE (full lines) and in the GCE (symbols).} 
\label{fig_npGGEvsGC}
\end{figure}

 \subsection{Comparing the GGE  with the grand canonical ensemble}\label{sec_GGEvsGCE}
A generic non-integrable system, with conserved number of particles, admits only $\hat N$ and $\hat H$ as 
local conserved charges. Therefore, if a large-time stationary state exists it should be described by the 
Grand Canonical Ensemble (GCE) 
\begin{equation}
\hat\rho_{GCE} = Z_{GCE}^{-1} \mathrm{e}^{-\mu_{GC}\hat{N}-\beta_{GC} \hat{H}},
\end{equation}
where, once again, the chemical potential $\mu_{GC}$ and the inverse temperature $\beta_{GC}$ are fixed by the 
initial conditions $\langle \hat N \rangle_0={\rm Tr}[\hat\rho_{GCE}\hat{N}]$, 
$\langle \hat H \rangle_0={\rm Tr}[\hat\rho_{GCE}\hat{H}]$.

It is worth investigating the qualitative and quantitative differences between the grand canonical ensemble  
and the GGE to provide testable predictions for experiments and numerical analyses. 
In our case, in order to estimate the differences in the expectation values of local observables, 
we compare the results previously obtained in the GGE with the grand canonical ensemble. In particular, the two Lagrange
multipliers $\mu_{GC}$ and $\beta_{GC}$ are fixed by the following set of equations (already in the TD limit)
\begin{eqnarray}
\int_{0}^{\infty} \frac{dp}{\pi}\frac{1}{1+\mathrm{e}^{\beta_{GC} p^2 + \mu_{GC}}} & = & 
\frac{n_{\mathcal{L}} + n_{\mathcal{R}}}{2} \\
\int_{0}^{\infty} \frac{dp}{\pi}\frac{p^2}{1+\mathrm{e}^{\beta_{GC} p^2 + \mu_{GC}}} & = &
\frac{\mathcal{E}_{\mathcal{L}} + \mathcal{E}_{\mathcal{R}}}{2},\nonumber
\end{eqnarray}
which can be rewritten in terms of polylogarithm functions ${\rm Li}_{n}(z)=\sum_{k=1}^{\infty}z^k / k^n$ as
\begin{eqnarray}\label{GCE_multipliers}
{\rm Li_{1/2}}(-{\rm e}^{-\mu_{GC}}) & = & 
-\beta_{GC}^{1/2} \sqrt{\pi}  (n_{\mathcal{L}} + n_{\mathcal{R}})\\
{\rm Li_{3/2}}(-{\rm e}^{-\mu_{GC}}) & = & 
-2 \beta_{GC}^{3/2} \sqrt{\pi}  (\mathcal{E}_{\mathcal{L}} + \mathcal{E}_{\mathcal{R}}),\nonumber
\end{eqnarray}
from which $\mu_{GC}$ is given by the solution of
\be
\frac{{\rm Li_{3/2}}(-{\rm e}^{-\mu_{GC}}) }{[{\rm Li_{1/2}}(-{\rm e}^{-\mu_{GC}})]^3} 
= \frac{2}{\pi} \frac{\mathcal{E}_{\mathcal{L}} + \mathcal{E}_{\mathcal{R}}}{(n_{\mathcal{L}} + n_{\mathcal{R}})^3},
\ee
and, therefore, $\beta_{GC}$ is obtained plugging the numerically found $\mu_{GC}$ in one of the two equations
in Eq.(\ref{GCE_multipliers}). In Figure \ref{fig_npGGEvsGC} (a) we compare the GGE mode occupation distribution
with the grand canonical results. Notice how the GCE gives completely different predictions with respect to the exact
GGE results. In particular, the discrepancy reduces as the two initial temperatures approach to each other. 
Actually, for $\beta_{\mathcal{L}} = \beta_{\mathcal{R}}$, the initial state is already a thermal state and, 
in this case, the dynamics is trivial.
 
Nonetheless, as far as short-range correlation function is concerned, the grand canonical prediction
coincides with the GGE prediction up to order $O(z^4)$. Indeed, expanding $C_{GGE}(z)$ around $z=0$
one has
\bea
C_{GGE}(z) & = & \int_{0}^{\infty} \frac{dp}{\pi} n(p) - \frac{z^2}{2}\int_{0}^{\infty} \frac{dp}{\pi} p^2 n(p) + O(z^4)\nonumber\\
 & = & \frac{n_{\mathcal{L}} + n_{\mathcal{R}}}{2}  -  
 \frac{\mathcal{E}_{\mathcal{L}} + \mathcal{E}_{\mathcal{R}}}{2} \frac{z^2}{2} + O(z^4),
\eea
which agrees with the $2^{nd}$-order expansion of the correlation function evaluated in the GCE.
In Figure \ref{fig_npGGEvsGC} (b) we compare, in $\log$-scale, the GGE and the GCE correlation function.

\section{Conlusions}\label{Concl}
In this paper we studied analytically and numerically the non-equilibrium dynamics of a Fermi gas 
initially prepared into two halves at different temperatures. After putting in contact the two halves, the system 
is left to evolve with a non-interacting Hamiltonian. 
In a first step, we considered an infinitely extend system and we fully characterized the dynamics of the
particles and energy profiles by means of a hydrodynamic approach \cite{akr08,cark12,wck13}. 
From those results we obtained the analytical expression for the particle and energy currents 
which perfectly matches the CFT predictions \cite{bd12}. 

Nevertheless, we argued that, the non-equilibrium stationary state describing such currents 
represents a regime which is completely destroyed whenever the system is finite.
Real systems are usually finite indeed, and the boundaries should play a crucial role in the 
equilibration mechanism.

Thus, we stressed that the mechanism which leads to the equilibration is due to the interference of the particles
going around the finite system many times \cite{csc13}. We prove that for long time and in a proper TD limit, 
i.e. taking into account the effects of the boundaries, any finite subsystem becomes truly stationary and its 
behavior is described by a GGE which only depends on the post-quench occupation mode distribution.  
This provides a proof of a GGE for an inhomogeneous initial state constructed from two halves at two different
temperatures.

\section{Acknowledgements}
The authors are grateful to Viktor Eisler and Giuseppe Mussardo for correspondence.
M. C. thanks Pasquale Calabrese for helpful discussions and acknowledges the 
ERC for financial support under Starting Grant 279391 EDEQS.



\end{document}